\documentclass[sigconf,nonacm]{acmart}

\newcommand\vldbdoi{XX.XX/XXX.XX}
\newcommand\vldbpages{XXX-XXX}
\newcommand\vldbvolume{18}
\newcommand\vldbissue{6}
\newcommand\vldbyear{2025}
\newcommand\vldbauthors{\authors}
\newcommand\vldbtitle{\shorttitle} 
\newcommand\vldbavailabilityurl{https://github.com/Spear-Neil/IndexResearch.git}
\newcommand\vldbpagestyle{empty} 

\usepackage{xcolor}
\usepackage{pifont}
\usepackage{listings}
\usepackage{multicol}
\usepackage{graphicx}
\usepackage{listings}
\usepackage{enumitem}
\usepackage{booktabs}
\usepackage{hyperref}
\usepackage{multicol}
\usepackage{tcolorbox}
\usepackage{algorithm}
\usepackage{mdframed}
\usepackage{algpseudocode}
\usepackage{bbding}
\usepackage{balance}

\definecolor{darkgreen}{rgb}{0.0, 0.4, 0.0}
\definecolor{darkblue}{rgb}{0.0, 0.0, 0.5}
\definecolor{darkpink}{rgb}{0.6, 0.0, 0.4}

\definecolor{revisioncolor}{rgb}{0.0, 0.0, 0.0}

\begin{document}
\title{FB$^+$-tree: A Memory-Optimized B$^+$-tree with Latch-Free Update}


\author{Yuan Chen}
\affiliation{
  \institution{School of Computer Science, Wuhan University}
}
\email{yuan.chen@whu.edu.cn}

\author{Ao Li}
\affiliation{
  \institution{School of Computer Science, Wuhan University}
}
\email{leo210@whu.edu.cn}

\author{Wenhai Li}
\affiliation{
  \institution{School of Computer Science, Wuhan University}
}
\email{lwh@whu.edu.cn}

\author{Lingfeng Deng}
\affiliation{
  \institution{School of Computer Science, Wuhan University}
}
\email{lingfengdeng@whu.edu.cn}






\begin{abstract} 
B$^+$-trees are prevalent in traditional database systems due to their versatility and balanced structure. While binary search is typically utilized for branch operations, it may lead to inefficient cache utilization in main-memory scenarios. In contrast, trie-based index structures drive branch operations through prefix matching. While these structures generally produce fewer cache misses and are thus increasingly popular, they may underperform in range scans because of frequent pointer chasing.


This paper proposes a new high-performance B$^+$-tree variant called \textbf{Feature B$^+$-tree (FB$^+$-tree)}. Similar to employing bit or byte for branch operation in tries, FB$^+$-tree progressively considers several bytes following the common prefix on each level of its inner nodes\textemdash referred to as features, which allows FB$^+$-tree to benefit from prefix skewness. FB$^+$-tree blurs the lines between B$^+$-trees and tries, while still retaining balance. In the best case, FB$^+$-tree almost becomes a trie, whereas in the worst case, it continues to function as a B$^+$-tree. Meanwhile, a crafted synchronization protocol that combines the link technique and optimistic lock is designed to support efficient concurrent index access. Distinctively, FB$^+$-tree leverages subtle atomic operations seamlessly coordinated with optimistic lock to facilitate latch-free updates, which can be easily extended to other structures. Intensive experiments on multiple workload-dataset combinations demonstrate that FB$^+$-tree shows comparable lookup performance to state-of-the-art trie-based indexes and outperforms popular B$^+$-trees by 2.3x$\ \sim\ $3.7x under 96 threads. FB$^+$-tree also exhibits significant potential on other workloads, especially update workloads under contention and scan workloads.

\end{abstract}

\maketitle

\pagestyle{\vldbpagestyle}
\begingroup\small\noindent\raggedright\textbf{PVLDB Reference Format:}\\
\vldbauthors. \vldbtitle. PVLDB, \vldbvolume(\vldbissue): \vldbpages, \vldbyear.\\
\href{https://doi.org/\vldbdoi}{doi:\vldbdoi}
\endgroup
\begingroup
\renewcommand\thefootnote{}\footnote{\noindent
$^{\ast}$Wenhai Li is the corresponding author.\\
This work is licensed under the Creative Commons BY-NC-ND 4.0 International License. Visit \url{https://creativecommons.org/licenses/by-nc-nd/4.0/} to view a copy of this license. For any use beyond those covered by this license, obtain permission by emailing \href{mailto:info@vldb.org}{info@vldb.org}. Copyright is held by the owner/author(s). Publication rights licensed to the VLDB Endowment. \\
\raggedright Proceedings of the VLDB Endowment, Vol. \vldbvolume, No. \vldbissue\ %
ISSN 2150-8097. \\
\href{https://doi.org/\vldbdoi}{doi:\vldbdoi} \\
}\addtocounter{footnote}{-1}\endgroup

\ifdefempty{\vldbavailabilityurl}{}{
\vspace{.3cm}
\begingroup\small\noindent\raggedright\textbf{PVLDB Artifact Availability:}\\
The source code, data, and/or other artifacts have been made available at \url{\vldbavailabilityurl}.
\endgroup
}

\section{introduction}
\label{section1}
The storage engine significantly influences the overall performance of a database system, especially the index data structure. Even in main-memory systems, index query operations account for $14\%\sim94\%$ of the overhead \cite{walker}. B-trees (B$^+$-tree, B$^*$-tree, etc.) are ubiquitous in disk-based database systems because of their prominent IO efficiency \cite{comer}, yet scarcely exist in main-memory database systems due to their poor utilization of hierarchical caches. ART \cite{art}, Masstree \cite{masstree}, and other trie-based structures are generally more efficient than main-memory B$^+$-tree \cite{art, masstree, hot, reducing, hat-trie}, which makes them more prevailing in modern main-memory systems (e.g., Silo \cite{silo} and HyPer \cite{hyper}). What makes main-memory B$^+$-tree untenable is the orders of magnitude difference in access latency between disk and memory. Generally, main memory provides an access latency of $80\sim100$ ns, whereas the disk and flash access latency are about 10 ms and 50 us, respectively \cite{pmem}. When IO dominates the overhead of a B$^+$-tree, the impact of CPU caches is nearly negligible. However, effective cache utilization becomes prominent in enhancing index performance when IO is eliminated \cite{csbtree, palm, caches, art, masstree}.

\begin{figure}
    \centering
    \includegraphics[width=0.48\textwidth]{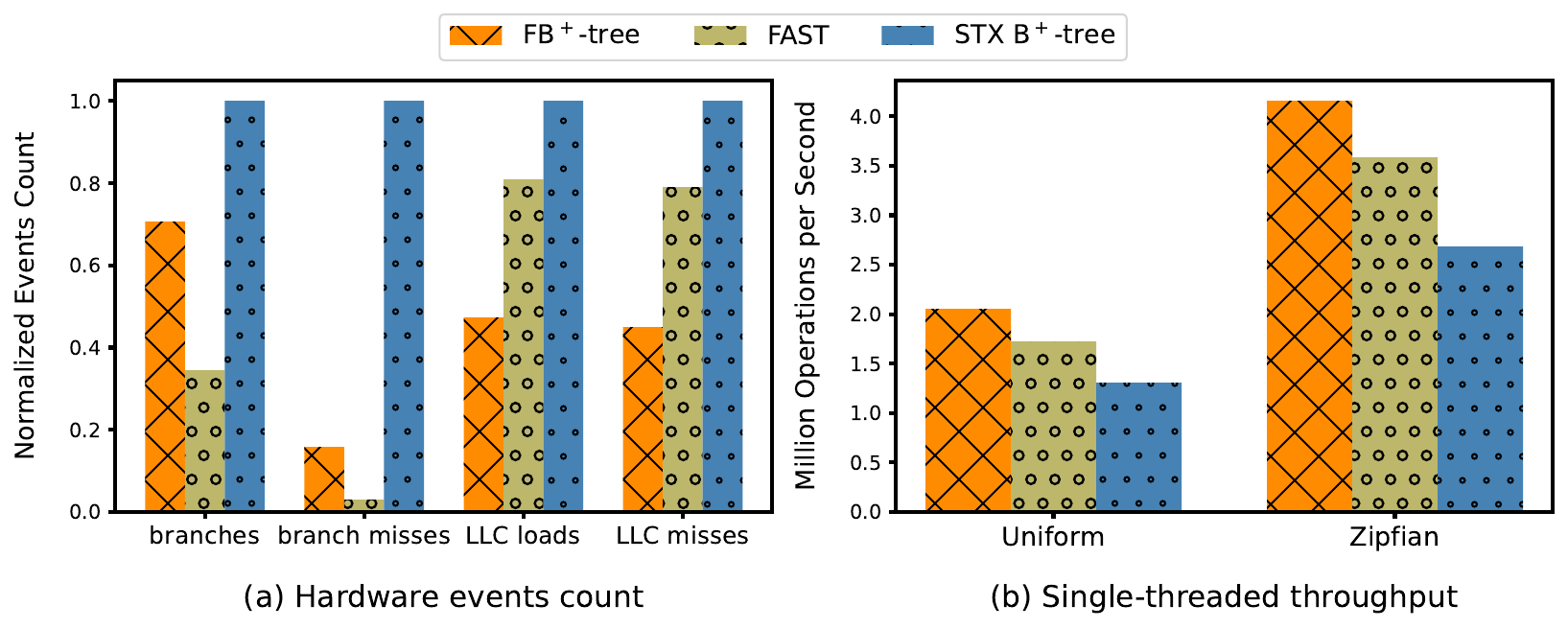}
    \captionsetup{skip=5pt}
    \caption{\textcolor{revisioncolor}{(a) Hardware statistics under uniform distribution, (b) Single-threaded throughput under different distributions.}}
    \label{fig:cache_miss}
    \vspace{-0.5em}
\end{figure}

\vspace{0.5em}
\textit{\textbf{Cache and Branch Optimization.}}
A B$^+$-tree consists of inner nodes and leaf nodes. Inner nodes serve as index nodes, containing anchor keys\footnote{Also known as separator keys, we follow the term anchor keys used in previous work.} and pointers to child nodes. Key-value pairs are stored in leaf nodes sequentially or semi-ordered \cite{fptree, lb+tree}. All leaf nodes are linked together in a linked list, enabling efficient range queries. A query starts from the root node and navigates to a leaf node using binary search for branch operations, i.e., index traversal. Thanks to such design, B$^+$-tree shows efficient disk access. But when it comes to main-memory scenarios, B$^+$-tree suffers from several interconnected and complex problems.

First, B$^+$-tree's structure and binary search render it inefficient on cache utilization, as detailed in Section \ref{section3}. Second, each comparison in binary search depends on the outcome of the previous comparison, and the result of each comparison is hard to predict \cite{art}. This intrinsic characteristic of binary search hinders the full exploitation of modern CPUs' computational and memory-level parallelism potential. Hard-to-predict comparisons pose challenges for dynamic branch prediction, causing modern CPUs' long pipelines to stall. Dependences between instructions render superscalar, speculation, and out-of-order execution powerless \cite{intel-dev-manual,intel-opt-manual}.

In this paper, we propose a new optimistic main-memory B$^+$-tree variant named \textbf{Feature B$^+$-tree (FB$^+$-tree)} to tackle these issues. FB$^+$-tree effectively mitigates the mismatch between B$^+$-tree's memory access patterns and the intrinsic characteristics of hardware architecture through progressively byte-wise parallel processing. In addition, FB$^+$-tree employs a simple for loop for branch operations in most cases. This enables FB$^+$-tree to fully leverage modern CPUs' parallelism potential and to facilitate sequential memory access. \textcolor{revisioncolor}{Figure \ref{fig:cache_miss} illustrates a comparison of throughput and hardware event statistics of FB$^+$-tree, FAST \cite{fast} and STX B$^+$-tree \cite{TLX} during 100 million random 64-bit integer lookups following uniform and zipfian distributions on a dual-socket, 48-core (96-hyperthread) server. Hardware events during lookups under uniform distribution are measured with \textit{perf}. Compared to STX B$^+$-tree, FB$^+$-tree shows fewer branch instructions, branch misses, LLC loads and LLC misses. The FAST tree is a binary tree that collapses multiple nodes into one large node to facilitate SIMD instructions and cache consciousness. Although FAST has fewer branch instructions and branch misses, FB$^+$-tree has better cache utilization due to its discriminative byte processing. This makes FB$^+$-tree outperform STX B$^+$-tree and FAST in single-threaded execution.}


\begin{figure}
    \centering
    \includegraphics[width=0.48\textwidth]{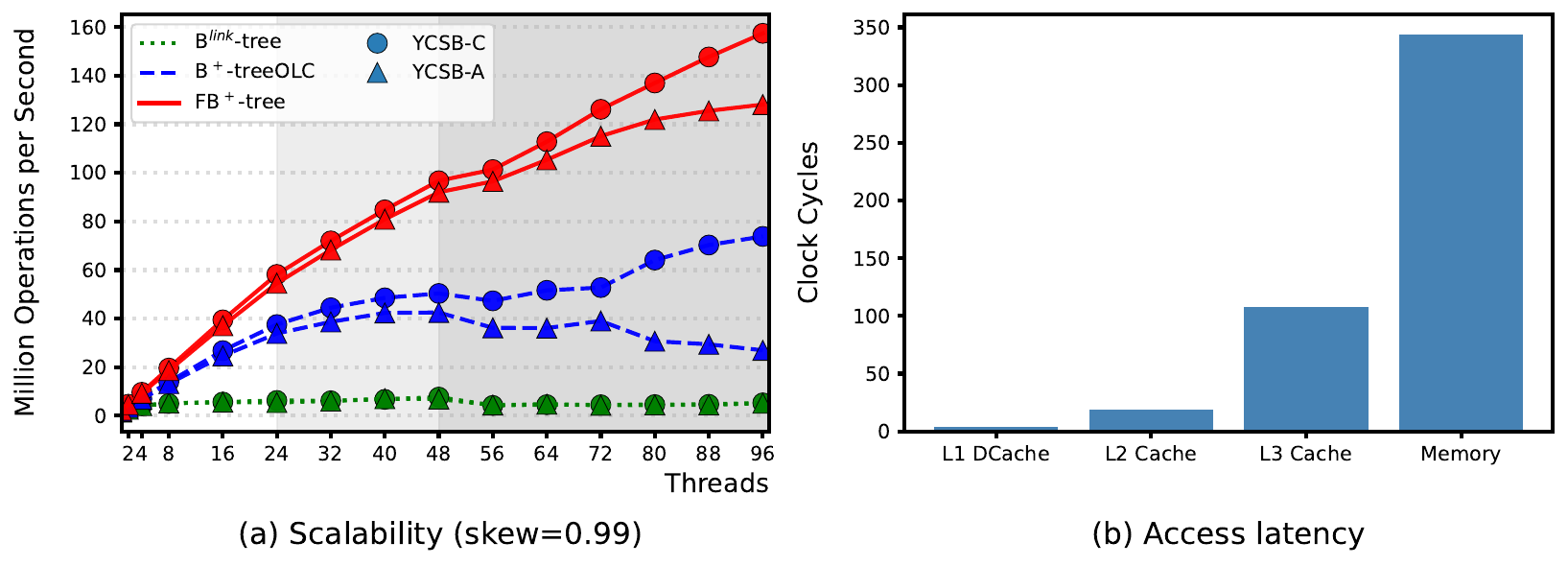}
    \captionsetup{skip=5pt}
    \caption{\textcolor{revisioncolor}{(a) Multi-core scalability of B$^{link}$-tree, B$^+$-treeOLC, and FB$^+$-tree, (b) Access latency of memory hierarchies.}}
    \label{fig:scalability_latency}
    \vspace{-0.5em}
\end{figure}

\vspace{0.5em}
\textit{\textbf{Synchronization Protocol.}}\label{sec:intro-sync}
Another crucial factor that impacts the overall performance of a database system is the synchronization protocol of its index. \textcolor{revisioncolor}{A fine-grained locking protocol, such as lock coupling (hand-over-hand lock) \cite{lock-coupling} and B$^{link}$-tree \cite{blink, symmetric}, has poor scalability in main-memory database systems. Optimistic lock utilizes a version combined with a lock to detect changes within a node \cite{olc, cscc, optik, masstree}.} These protocols enable latch-free\footnote{We follow the term "latch-free" used in OLFIT \cite{cscc} which means without acquiring locks to distinguish it from the term "lock-free" which means non-blocking. We follow the synchronization literature to use "lock" except "latch-free".} index traversal except during node modification, thereby ensuring high scalability. FB$^+$-tree employs a similar yet highly optimized protocol for index traversal and incorporates the link technique \cite{blink} for concurrent structure modification (i.e., node split and merge). 

\textcolor{revisioncolor}{The salient innovation of FB$^+$-tree's synchronization protocol is latch-free update. Previous work acquires a lock on a node or locks on more nodes when performing an update. They protect overmuch auxiliary computations in critical sections and frequently retry if they cannot hold the lock, leading to coherence cache invalidations. Such locks thus prevent other threads from performing even unrelated operations and make updates poorly scalable under high-contention workloads. As shown in Figure \ref{fig:scalability_latency}(a), B$^+$-treeOLC synchronized by optimistic lock coupling \cite{olc} is more scalable than B$^{link}$-tree on YCSB-C workload (read-only). However, it suffers from performance collapse on YCSB-A workload (50\%-read, 50\%-update). FB$^+$-tree mitigates such collapse via latch-free update.}

In short, FB$^+$-tree employs a more fine-grained synchronization protocol which enables an update to be performed without acquiring any locks. FB$^+$-tree executes an update using the compare-and-swap (CAS) instruction. This way, a lookup can be executed simultaneously with updates, and updates only contend on the same key-value pairs. Our protocol offers a general technique for performing updates without acquiring any locks. Index structures synchronized by optimistic lock can implement latch-free updates with a few changes to existing code. Benchmarks on YCSB workload A (update heavy) demonstrate that FB$^+$-tree outperforms existing indexes with optimistic lock under contentions.

\vspace{0.5em}
\textit{\textbf{Contributions and Paper Organization.}}
This paper makes the following contributions: \par
\begin{itemize}[leftmargin=*]
    \item \textcolor{revisioncolor}{A technique supporting arbitrary key types named feature comparison is introduced, which allows FB$^+$-tree to work like a trie and leverage modern CPUs' computational and memory-level parallelism potential. This technique combined with hashtags in leaf nodes enables FB$^+$-tree to outperform typical B$^+$-trees in both performance and scalability on index traversal.}
    
    \item \textcolor{revisioncolor}{Thanks to feature comparison, FB$^+$-tree does not need to frequently access anchor keys like typical B$^+$-trees. FB$^+$-tree thus only stores pointers to anchor keys for space efficiency.}
    
    \item A highly optimized optimistic synchronization protocol is devised to facilitate multi-core scalable concurrent index accesses. In particular, the protocol introduces a general latch-free update technique that utilizes CAS primitive for update operations without holding any locks. 
    
    \item Comprehensive experiments are conducted on multiple workload-dataset combinations for existing index structures and FB$^+$-tree. The results demonstrate that FB$^+$-tree outperforms popular B$^+$-trees by 2.3x$\ \sim\ $3.7x on read-only workloads under 96 threads. On update-heavy workloads, FB$^+$-tree shows the best scalability thanks to latch-free update technique.
\end{itemize}
The rest of this paper is organized as follows. Section \ref{section2} introduces background and related work. Section \ref{section3} presents the motivation, data structure, and algorithm of FB$^+$-tree. Section \ref{section4} describes the synchronization protocol enabling FB$^+$-tree multi-core scalable. Section \ref{section5} shows the experiment setup, workloads, and evaluation results. The conclusion is given in Section \ref{section6}.
\section{related work}
\label{section2}
Querying an index for a key (i.e., index traversal) involves the process of narrowing the search key space until confirming whether the key exists. In B-trees, the process relies on comparisons between the target key and the anchors. Meanwhile, in tries, it depends on prefix matching. This section provides some background and related work on ordered index. In addition, previous work on synchronization protocols is presented at the end.

\vspace{0.5em}
\textit{\textbf{Memory Access.}} From a more detailed perspective, index traversal can be divided into memory access and actual computation. Compared to memory access, the computational overhead is almost negligible. 
On modern machines, several to hundreds of instructions can be evaluated and retired in parallel, and most SIMD instructions have 1-cycle throughput \cite{primer, intel-dev-manual, intel-opt-manual}. However, memory access typically takes several hundred cycles to complete, as shown in Figure \ref{fig:scalability_latency}(b). Various main-memory ordered indexes have been proposed to optimize cache utilization. We categorize existing ordered indexes into three groups: B-trees, tries, and learned indexes. Next, we introduce their related work respectively.

\vspace{0.5em}
\textcolor{revisioncolor}{\textit{\textbf{B-trees.}} B-trees are a class of balanced, comparison-based indexes\footnote{Skip list \cite{skip-list} and its variants are also comparison-based.}, encompassing many variants such as B-tree, B$^*$-tree, B$^+$-tree, binary B-tree, AVL tree, and others \cite{comer}. As memory capacity increased, the T-tree placed multiple records in one binary tree's node for main-memory database systems \cite{Ttree}. The CSS-trees \cite{csstree} and CSB$^+$-tree \cite{csbtree} store each node's children in contiguous memory to mitigate pointer chasing. Bender et al. conducted an intensive theoretical analysis and experimental evaluation on cache-oblivious B-trees \cite{cob, cosb, ccob, performance}. Chen et al. proposed to optimize B$^+$-trees' performance through prefetching for both cache and disk \cite{chen1, chen2}.}

\textcolor{revisioncolor}{Currently, software prefetching is extensively used in main-memory indexes \cite{masstree, art, hot}, as it targets short array streams and irregular memory address patterns \cite{when, analysis, guided}. The prefix B-trees explored not directly storing keys but constructing prefixes in inner nodes \cite{prefixbtree, foster}. The pkT-trees and pkB-trees store fixed-size parts of keys directly in the tree nodes
\cite{partial1, partial2}. The B$^2$-tree organizes inner nodes as embedded trie-like structures for indexing string keys \cite{b2tree}. The DB$^+$-tree \cite{dbtree} incorporates the discriminative bits from HOT \cite{hot} into B-trees. From another perspective, k-Ary, FAST
, P-ary, etc. exploit binary search with SIMD instructions to leverage parallelism of both CPUs and GPUs \cite{k-ary, p-ary, fast, hybrid-btree, boosting}. PALM performs multiple concurrent queries in batches \cite{palm}.}

\begin{figure}[t]
    \centering
    \includegraphics[width=0.45\textwidth]{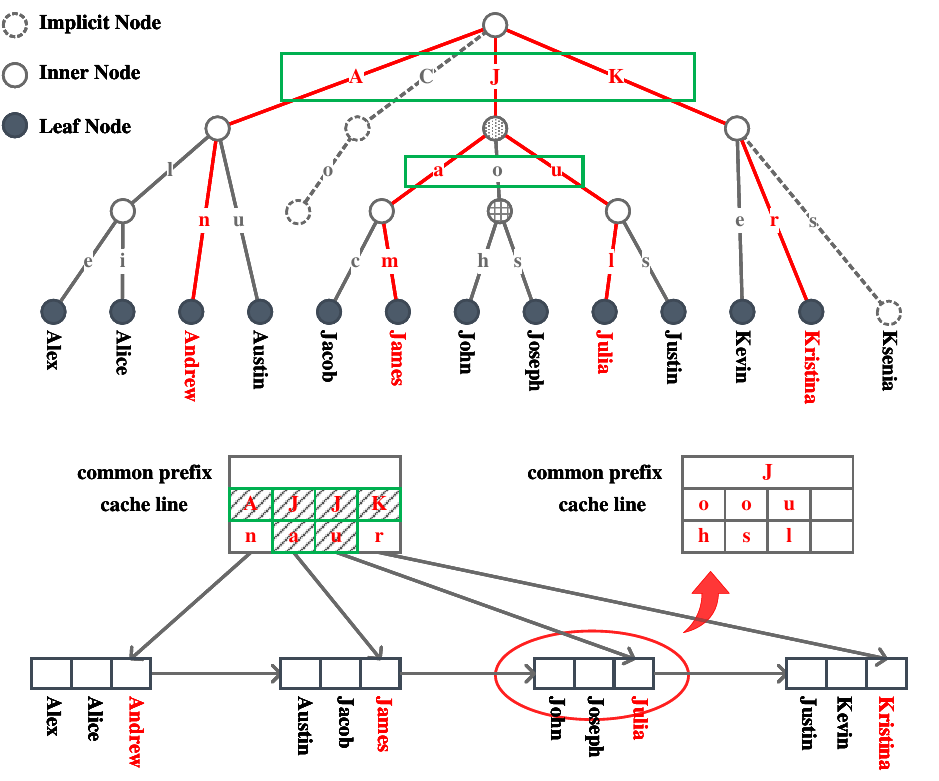}
    \captionsetup{skip=5pt}
    \caption{Illustration of a trie and the main idea of FB$^+$-tree}
    \label{fig:idea}
    \vspace{-0.5em}
\end{figure}

\vspace{0.5em}
\textcolor{revisioncolor}{\textit{\textbf{Tries.}} Tries, also known as radix trees, prefix trees, or digital search trees, drive branch operations through prefix matching, as illustrated in Figure \ref{fig:idea}. These data structures directly use the digital representation of keys and may have excessive worst-case space consumption \cite{art}. The Patricia trie introduces path compression and only stores prefixes of keys in trie nodes \cite{patricia}. The burst trie substitutes leaf nodes with containers maintaining a small set of keys \cite{burst-trie}. The HAT-trie maintains hash tables as containers for better performance \cite{hat-trie}. The generalized prefix tree is a trie with variable prefix length for indexing arbitrary data types \cite{generalized-trie}.}

\textcolor{revisioncolor}{The Adaptive Radix Tree (ART) adaptively uses four different node layouts depending on the number of child nodes \cite{art}. Path compression and lazy expansion allow ART to efficiently index string keys by collapsing nodes. The Masstree is a trie-like concatenation of B$^+$-trees, where each trie node is a B$^+$-tree indexing different 8-byte slices of keys \cite{masstree}. The Height Optimized Trie (HOT) dynamically varies the number of discriminative bits considered at each node \cite{hot, full-hot}. Similar to B$^+$-trees, another interesting design Wormhole \cite{wormhole} maintains a double-linked list of leaf nodes. It then constructs a trie with all prefixes of the lower bound of leaf nodes and represents the trie with a hash table. The Cuckoo Trie shares a similar idea and exploits memory-level parallelism to alleviate pointer chasing \cite{cuckoo-trie}.}

\vspace{0.5em}
\textit{\textbf{Learned Indexes.}} Indexes can be viewed as models to map a key to the position of a record. Kraska et al. proposed learned indexes (machine learning models) as replacements for index structures. They have demonstrated that learned indexes have the potential to offer benefits over state-of-the-art indexes \cite{learned-index}. The ALEX exploits linear regression combined with exponential search to guarantee accuracy and enable adaptive update \cite{alex, apex}. Some previous work has explored indexing string keys using learned indexes \cite{last-mile, radixspline, accelerating, lits}. Learned indexes and database systems with AI have become a prominent research topic, and extensive work has been dedicated to making them practical \cite{oasis, accelerating, pilotscope,learned-disk, hyper-learned, ai-db}.

\textcolor{revisioncolor}{In summary, previous research has demonstrated that tries outperform B-trees in terms of point lookup. For range iteration, B-trees, especially B$^+$-trees, demonstrate better performance because of their balanced structure. The Wormhole combines the strengths of B$^+$-tree and trie. Unfortunately, its hashed representation of meta-trie and indirect ordered leaf node may incur significant overhead for insert and range scan operations, respectively. Learned indexes may offer advantages; however, challenges still exist in achieving accuracy, adaptive update capabilities, and indexing string keys.}

\vspace{0.5em}
\textcolor{revisioncolor}{\textit{\textbf{Synchronization on Indexes.}} Designing an efficient synchronization protocol is critical and challenging. Lock coupling \cite{lock-coupling} and B$^{link}$-tree \cite{blink} lock only one node simultaneously, thus demonstrating good scalability in traditional database systems. Optimistic lock coupling, OLFIT, Masstree, and many other indexes combine version with lock to avoid coherence cache miss \cite{olc, optik, cscc, masstree, art, hot, cuckoo-trie}. In these protocols, readers verify the version to detect changes within a node without acquiring the lock. Meanwhile, writers acquire a write lock and update the version when modifying a node. The Read-Optimized Write Exclusion (ROWEX) protocol used by ART and HOT goes one step further by only providing exclusion relative between writers while allowing readers to always succeed without block nor restart \cite{olc, rowex, hot}.}

OLFIT and several indexes employ a bottom-up strategy for structure modification, e.g. split and merge, and adopt the link technique and high key from B$^{link}$-tree for concurrency support \cite{cscc, masstree}. 
Optimistic lock coupling provides a more general approach to synchronize index structures \cite{olc, rowex}, utilizing a top-down strategy for concurrent structure modification. If a change occurs during index traversal, it restarts from the root. ROWEX utilizes subtle atomic operations to ensure that reads are always consistent and never restart \cite{rowex}. OptiQL incorporates queue-based locking and opportunistic read techniques into optimistic lock coupling to mitigate performance collapse under high contention \cite{optiql}. The Bw-tree implements lock-free operations through delta records and mapping table \cite{bw-tree, openbw}, but it may incur heavy overhead. Some work has proposed universal construction strategies to transform trees from sequential code to linearizable concurrent versions \cite{occualizer}.

We have learned lessons from previous work. FB$^+$-tree considers a B$^+$-tree from trie's prefix matching perspective. Common prefixes and parts of anchors are directly stored in inner nodes but with a quite different arrangement. Unlike k-ary, FAST, etc., which only consider data types supported by SIMD instructions, FB$^+$-tree facilitates byte-wise parallel processing for branch operations, allowing arbitrary data types. For concurrency support, FB$^+$-tree uses a highly optimized optimistic synchronization protocol and employs the link technique for concurrent structure modification. Furthermore, FB$^+$-tree allows multiple writers to perform updates on one leaf node without blocking other updates nor reads.

\section{The FB$^+$-tree data structure}
\label{section3}
In this section, we start by providing a comprehensive analysis of the overhead associated with comparison operations in B$^+$-trees. Next, we introduce our feature comparison technique, FB$^+$-tree's data structure, and the branch algorithm. The synchronization protocol is discussed in Section \ref{section4}.

\subsection{Dissonance between B$^+$-tree and Hardware}\label{dissonance}
As previously mentioned, binary search is employed to retrieve child nodes in index traversal of B$^+$-trees. In binary search, the anchor key for comparison is the median of the current search key space. This leads to a complex memory access pattern that underutilizes effective hardware and software prefetching. Moreover, hard-to-predict comparisons and dependences between comparisons make memory-level parallelism unattainable \cite{cuckoo-trie}.

Consider a B$^+$-tree indexing string keys, where only pointers to anchor keys are stored in inner nodes and the contents of anchors are stored in memory allocated via malloc. Assuming a fanout factor 256, binary search in an inner node would typically involve eight comparisons. On modern 64-bit machines, storing 256 pointers to anchors would require 256 * 8 bytes of space, equivalent to 32 cache lines. Consequently, each binary search would access six cache lines for pointers alone, whereas only eight pointers are effectively used. For B$^+$-trees indexing integer keys, similar issues arise. Several indexes therefore configure their nodes as 256 bytes with a proper fanout to alleviate this cache inefficiency. \cite{masstree, art, hot, openbw}.

After obtaining the pointer to an anchor, the content of the anchor is compared with that of the target key following dereference. No matter how many bytes the anchor has and how many bytes are used for comparison, at least one complete cache line is loaded from memory. However, in many cases, only a few bytes are necessary to establish the relative order between string keys (similarly for integer keys) \cite{prefixbtree, partial1, partial2}. Consequently, this results in wasted memory bandwidth and the eviction of several hot cache lines.

This problem is exacerbated on modern CPUs, because continuous several cache lines will be automatically prefetched but remain unused, leading to bandwidth waste \cite{intel-opt-manual, analysis}. In concurrent environments, these random small memory accesses further impede full utilization of memory bandwidth and Ultra Path interconnect (UPI) bandwidth \cite{cuckoo-trie, intel-opt-manual}. These problems result in B$^+$-trees' sub-optimal scalability even on read-only workloads.

\subsection{Feature Comparison}\label{feature-cmp}
\subsubsection{Motivation} 
\textcolor{revisioncolor}{Tries have advantages over B$^+$-trees in cache utilization, as they drive branch operations through prefix matching. The key idea of FB$^+$-tree is to integrate such mechanism into B$^+$-trees to benefit from architecture while preserving B$^+$-trees' properties, particularly balance. We start with the intrinsic similarities between B$^+$-trees and tries, as illustrated in Figure \ref{fig:idea}.}

\textcolor{revisioncolor}{In B$^+$-trees, the entire key space is divided into intervals defined by anchors in inner nodes, which serve as the upper and lower bounds of these intervals. In contrast, tries partition the entire key space into sub-trees using prefixes. B$^+$-trees and tries are functionally equivalent in this partition sense. In other words, trie nodes inherently imply the relative order between keys through their digital representation. Tries accomplish key space partitioning with byte-wise prefix matching (or using smaller radix), whereas comparisons between strings are also performed in a byte-wise manner. Comparison between binary keys can be conducted similarly after code transition, as detailed in Subsection \ref{optimization}. These similarities suggest that branch operation in B$^+$-trees could potentially be implemented similarly to prefix matching in tries.}

\textcolor{revisioncolor}{As shown in Figure \ref{fig:idea}, we build a trie and a B$^+$-tree for a collection of keys, where the anchor keys of the B$^+$-tree are highlighted in both structures. Consider a use case of querying the trie for the key "John". The most significant byte "J" is used to retrieve the dotted node following the root node. The search key space is then reduced to the sub-tree with the dotted node as root. Next, the latticed node is retrieved via the second byte "o", followed by reaching the leaf node "John". Revisiting this byte-wise prefix matching from a B$^+$-tree's comparison perspective, it implies the first byte "J" is compared with the first byte of the four anchor keys "Andrew", "James", "Julia", and "Kristina" simultaneously. After this byte-wise parallel comparison, the search key space is narrowed down to the keys greater than "Andrew" and less than "Kristina".}

\subsubsection{Algorithm}
\textcolor{revisioncolor}{Obviously, this byte-wise parallel comparison can be implemented with SIMD instructions. B$^+$-tree's branch operation can be implemented with progressively byte-wise processing, as shown in Figure \ref{fig:idea}. As anchor keys may share a common prefix, byte-wise processing on the common prefix is unproductive and space-inefficient. FB$^+$-tree constructs the common prefix in inner nodes and performs byte-wise parallel processing on the following bytes. \textit{An interesting property similar to prefixes in tries is that the common prefix length of a child node is no smaller than that of its parent node.} Branch operation could skip the common prefix and index traversal could gradually handle different slices of a target key on each level of inner node. FB$^+$-tree thus almost becomes a trie in terms of computational complexity.}

\textcolor{revisioncolor}{One reason why tries are generally more cache-conscious is their capability to fit complex data distribution or prefix skewness. Trie nodes shared by more keys would reside in a cache level closer to CPUs. By contrast, B$^+$-trees mechanically select the middle key as the anchor key during node split, which is unconscious of prefix skewness. In other words, several anchors may share a much longer common prefix (sparse keys). Performing byte-wise parallel processing over all the remaining bytes following the common prefix thus may be inefficient. For space efficiency, FB$^+$-tree only stores a few fixed-size discriminative bytes of anchors directly in inner node. A binary search over suffixes will be performed when byte-wise parallel comparison cannot achieve branch operation.}

\begin{figure}[t]
    \centering
    \includegraphics[width=0.48\textwidth]{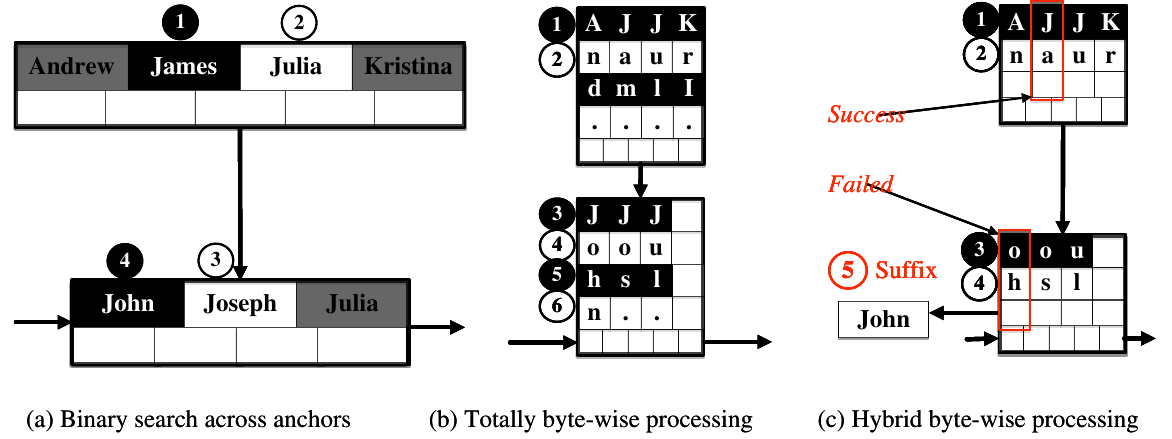}
    \captionsetup{skip=5pt}
    \caption{\textcolor{revisioncolor}{Binary search vs. FB$^+$-tree's branch algorithm.}}
    \label{fig:comp}
    \vspace{-0.5em}
\end{figure}

\textcolor{revisioncolor}{For example, consider the B$^+$-tree's third leaf node shown in Figure \ref{fig:idea} as an inner node. We show the process of querying the key "John" with binary search, totally byte-wise parallel processing, and FB$^+$-tee's hybrid branch algorithm respectively in Figure \ref{fig:comp}, where the numbers highlight the order of memory access along hierarchical inner nodes. Binary search generates irregular and small memory access, as shown in Figure \ref{fig:comp}(a). Totally byte-wise parallel processing does not take into account common prefixes and sparse keys, as shown in Figure \ref{fig:comp}(b). Supposing only two bytes following common prefix are stored directly in inner nodes in Figure \ref{fig:comp}(c). In the root node, branch operation succeeds after two byte-wise comparisons. In the child node, the common prefix "J" can be skipped. In this special case, byte-wise parallel processing fails in branch and a binary search over suffixes has to be performed.}

\subsubsection{Discussion}
\textcolor{revisioncolor}{With this byte-wise parallel processing on discriminative bytes, FB$^+$-tree works like a trie and alleviates direct access to anchors. As index traversal descends to a leaf node, common prefix length grows and branch operation gradually processes different slices of a target key. Therefore, binary search over suffixes would occur with very low probability. FB$^+$-tree thus evolves into a trie in the best case. In the worst case, branch operations on all inner nodes depend on binary search over suffixes, which is almost impossible. FB$^+$-tree thus degenerates into a B$^+$-tree.}

\textcolor{revisioncolor}{In most cases, FB$^+$-tree forms a hybrid structure that combines the characteristics of both B$^+$-trees and tries. Prefix matching is significantly efficient for dense keys but sub-optimal for sparse keys\footnote{For tries, node collapse or path compression can mitigate this problem but not completely. HOT dynamically changes the number of discriminative bits used in a node, and meanwhile almost stays a constant fanout. HOT thus achieves almost extreme cache utilization for lookup. Unfortunately, the common problems of trie structures still exist, which is inefficient and complicated concurrent range iteration.} \cite{art, hot, masstree}. Binary search generally incurs expensive cache misses, while not affected by the sparseness of keys. FB$^+$-tree employs byte-wise parallel processing for dense keys to enhance cache consciousness. For sparse keys, this byte-wise processing on discriminative bytes acts as a filter to narrow the search space for binary search. This hybrid structure allows FB$^+$-tree to efficiently index both sparse and dense keys while maintaining a balanced structure that facilitates efficient range iteration.}

\textcolor{revisioncolor}{To some extent, these discriminative bytes imply some data distribution characteristics of the keys in node intervals (sparse or dense). Therefore, we refer to these bytes as features and name this byte-wise parallel processing as feature comparison. Even though FB$^+$-tree stores feature bytes, FB$^+$-tree only stores pointers to anchor keys in inner nodes, making it even more space-efficient than typical B$^+$-trees. Next, we introduce the node implementation, algorithm, and some optimizations.}

\begin{figure}[t]
\vspace{-0.5em}
\centering
\Small
\begin{tcolorbox}[colback=white!100,colframe=black!100,boxrule=0.8pt,arc=0pt]
\vspace{-1em}
\begin{lstlisting}[language=C++,basicstyle=\fontsize{6.2}{6.2}\ttfamily,columns=flexible,numbers=left,numbersep=5pt,tabsize=2,escapeinside=``,keywordstyle=\color{black},morekeywords={Control, node, String, KeyType, ValueType, uint64_t, KVPair,LeafNode},commentstyle=\color{black!50!white},numberblanklines=false]
  // for binary keys             // for string keys 
  struct `\textbf{InnerNode}`:             struct `\textbf{InnerNode}`:                       
      Control control;               Control control;
      int knum;                      int knum;       
      int plen;                      int plen;    
      char prefix[8];                String* huge;    // huge prefix
      node* next;                    node* next;           
      char features[fs][ns];         char features[fs][ns];   
      node* children[ns];            char tiny[224];  // embedded prefix  
                                     String* anchors[ns];
                                     node* children[ns];
                                   
  struct `\textbf{LeafNode}`:              struct `\textbf{String}`:
      Control control;               int len;             
      uint64_t bitmap;               char str[];          
      KeyType high_key;                        
      LeafNode* sibling;         struct `\textbf{KVPair}`:
      char tags[ns];                 KeyType key;         
      KVPair* kvs[ns];               ValueType value;     
\end{lstlisting}
\vspace{-1em}
\end{tcolorbox}
\captionsetup{skip=5pt}
\caption{Node structures of FB$^+$-tree.}
\label{fig:structure}
\vspace{-0.5em}
\end{figure}

\subsection{Node Implementation}\label{node-imp}
\noindent 
Figure \ref{fig:structure} illustrates the node structures of FB$^+$-tree. The \textit{control} is an 8-byte atomic variable that governs the synchronization behaviors. The \textit{knum} and \textit{plen} indicate the number of anchor keys and the length of the common prefix in an inner node. The \textit{next} specifies an inner node's sibling or last child. The \textit{fs} and \textit{ns} which can be manually configured represent the feature and node size, respectively. In leaf nodes, the \textit{bitmap} indicates whether the corresponding slot in \textit{kvs} is occupied and the \textit{tags} field contains the corresponding hashtags. The \textit{high\_key} is the upper bound of a leaf node.

Essentially, both structures of FB$^+$-tree's inner nodes and leaf nodes are identical to that of a typical B$^+$-tree, except for the features in inner nodes and the hashtags in leaf nodes. To facilitate concurrent structure modification, all nodes on the same level are linked in a single-linked list. Each node starts with an 8-byte \textit{control} field, indicating the node type. Binary keys and string keys share similar inner node and leaf node structures. 

\vspace{0.3em}
\textit{Leaf Node.} Typically, key-value pairs are ordered in leaf nodes, which could lead to time-consuming rearrangement and binary search during insertion. These complex operations within a critical section may limit multi-core scalability. To address these issues, we store the unsorted key-value pairs in a pointer array \textit{kvs} and utilize hashtags for efficient lookup, as in previous work \cite{fptree, lb+tree, wormhole}. Additionally, each leaf node includes a \textit{high\_key} to support concurrent structure modification, which indicates the upper bound. For binary keys, the \textit{high\_key} is directly stored in the node; for string keys, each leaf node maintains a pointer to \textit{high\_key}.

\begin{figure}[t]
\centering
\Small
\begin{tcolorbox}[colback=white!100,colframe=black!100,boxrule=0.8pt,arc=0pt]
\vspace{-1em}
\begin{lstlisting}[language=C++,keywordstyle=\color{black},morekeywords={String, uint64_t, KVPair, __m512i},emph={fs, features, plen, knum, children, tags, bitmap, kvs},emphstyle={\color{darkgreen}},commentstyle=\color{black!50!white},basicstyle=\fontsize{6.2}{6.2}\ttfamily,columns=flexible,numbers=left,numbersep=5pt,tabsize=2,escapeinside=``,numberblanklines=false]
  void* branch(String& key) { // inner node `\label{branch-begin}`
    void* node = nullptr;
    int pcmp = prefix_compare(key); `\label{lookup:prefix}`
    if (pcmp == 0) { // prefix matches
      int idx, fid, cmps = min(fs, key.len - plen);
      uint64_t mask, eqmask = (0x01ul << knum) - 1;
      for (fid = 0; fid < cmps; fid++) { `\label{fbegin}`
        char byte = key.str[plen + fid] + 128;  `\label{lookup:byte1}`
        `\colorbox{cyan!15}{mask = compare\_equal({\color{darkgreen}{features}}[fid], byte);}` `\label{lookup:cmp_eq}`
        mask = mask & eqmask;
        if (mask == 0) break;
        eqmask = mask;
      } // feature comparison `\label{fend}`
      if (fid < cmps) { 
        char byte = key.str[plen + fid] + 128; `\label{lookup:byte2}`
        `\colorbox{cyan!15}{mask = compare\_less({\color{darkgreen}{features}}[fid], byte);}` `\label{lookup:cmp_les}`
        mask = mask & eqmask;
        if (mask == 0) idx = index_least1(eqmask); `\label{lookup:indexleast1}`
        else idx = 64 - countl_zero(mask); `\label{lookup:countlzero}`  
      } else { // binary search on suffixes
        int hid = 64 - countl_zero(eqmask);
        int lid = index_least1(eqmask);
        idx = suffix_bs(key, plen + cmps, lid, hid); `\label{lookup:suf}`
      }
      node = children[idx];
    } else { node = children[0]; }
    return node;
  } `\label{branch-end}`

  KVPair* lookup(String& key) { // leaf node `\label{leaf-begin}`
    char tag = hash(key.str, key.len);
    uint64_t mask = compare_equal(tags, tag);  `\label{lookup:hash}`
    mask = mask & bitmap; // candidates
    while (mask) {
      int idx = index_least1(mask);
      KVPair* kv = kvs[idx].load(); `\label{up-begin}`
      if (kv != nullptr && key == kv->key)  `\label{cmp-content}`
        return kv; // key found `\label{up-end}`
      mask &= ~(0x01ul << idx);
    }
    return nullptr; // key not found
  } `\label{leaf-end}`

  // compare 64 bytes to a char 'c', AVX512
  uint64_t compare_equal(void* p, char c) { `\label{cmpeq-begin}`
    __m512i v1 = _mm512_loadu_si512(p);
    __m512i v2 = _mm512_set1_epi8(c);
    return _mm512_cmpeq_epi8_mask(v1, v2);
  } `\label{cmpeq-end}`
\end{lstlisting}
\vspace{-1em}
\end{tcolorbox}
\captionsetup{skip=5pt}
\caption{Lookup Related Algorithms}\label{lst:lookup}
\vspace{-1.5em}
\end{figure}

\vspace{0.3em}
\textit{Inner Node.} The anchor keys in an inner node are ordered. For binary keys, all bytes of an anchor are directly stored in \textit{features}. The common prefix is stored in \textit{prefix}. For better performance, the prefixes of anchors are truncated and the remaining bytes in \textit{features} are shifted adjacent to the \textit{next} field. 

For string keys, anchor keys are stored in a pointer array\textemdash\textit{anchors}. Thanks to feature comparison, branch operation would rarely access anchors. Therefore, unlike typical B$^+$-trees that copy anchor keys into inner nodes, FB$^+$-tree stores the actual contents of anchor keys in leaf nodes (i.e., \textit{high\_key}), while inner nodes only maintain pointers to \textit{high\_key}, which makes FB$^+$-tree more space-efficient. Whenever possible, the entire common prefix is embedded in the \textit{tiny} field\footnote{For the sake of concurrency support, the entire common prefix is embedded in an inner node, even if the parent and child nodes have exactly the same common prefix.}. Slab memory allocators, such as jemalloc and tcmalloc, always allocate a memory block not smaller than the required size. Therefore, we configure the size of \textit{tiny} to fully utilize any excess memory available. The \textit{huge} field points to the first anchor as the common prefix in case it is too long to reside in \textit{tiny}.

\subsection{Lookup and Update}
\noindent
Figure \ref{lst:lookup} presents the slightly simplified code for \textit{branch} in inner nodes and \textit{lookup} in leaf nodes, without considering concurrency. The lookup process in an FB$^+$-tree is identical to that in a typical B$^+$-tree. Index traversal utilizes the \textit{branch} algorithm (lines \ref{branch-begin}$\ \sim \ $\ref{branch-end}) to retrieve child nodes until a leaf node. Subsequently, the hash-based \textit{lookup} algorithm (lines \ref{leaf-begin}$\ \sim \ $\ref{leaf-end}) is executed to retrieve the key-value pair in the leaf node. Similar processes are used for binary keys. The update process of FB$^+$-tree has a minor difference from the lookup process (lines \ref{up-begin}$\ \sim \ $\ref{up-end}). An atomic update operation based on CAS is employed without holding any locks.

The \textit{branch} algorithm begins by comparing the target key with the common prefix (line \ref{lookup:prefix}). The most significant difference between FB$^+$-tree's branch algorithm and typical binary search lies in the feature comparison (lines \ref{fbegin}$\ \sim \ $\ref{fend}). Each byte of the target key following the prefix is compared with the corresponding byte in features (line \ref{lookup:cmp_eq}), until either the last feature is reached or there is no more matching byte. No matching bytes in features means that the child node could be determined immediately using the \textit{compare\_less} function (line \ref{lookup:cmp_les}). In cases where a binary search on suffixes is necessary, the bytes including both the common prefix and features are truncated to improve performance (line \ref{lookup:suf}). The reason why 128 is added to each byte (lines \ref{lookup:byte1} and \ref{lookup:byte2}) will be presented in Subsection \ref{optimization}.

The \textit{lookup} algorithm employs a hash fingerprint in leaf nodes. Candidates are first filtered with hashtags (line \ref{lookup:hash}), followed by a verification comparison using the real content to prevent false positives (line \ref{cmp-content}). Lines \ref{cmpeq-begin}$\ \sim \ $\ref{cmpeq-end} show the implementation of \textit{compare\_equal} using AVX512. BMI and SIMD instructions, such as LZCNT, are utilized for efficient bit manipulation, for instance, \textit{index\_least1} (line \ref{lookup:indexleast1}) and \textit{countl\_zero} (line \ref{lookup:countlzero}).

\subsection{Insert and Remove}
A key-value insertion into a leaf node simply requires locating an empty slot using hashtags and then installing the key-value into \textit{kvs}. FB$^+$-tree adopts a bottom-up strategy for insertion that involves structure modification. The algorithm for finding the position to insert an anchor key into an inner node also relies on feature comparison. The primary disparity lies in the recomputation of the common prefix and features. The common prefix of an inner node is recomputed only when the new anchor is less than the minimum anchor key. In most cases, an anchor key insertion could be easily conducted by inserting the pointer and features. Meanwhile, inner nodes are modified only during structure modifications. Remove follows a similar process.

\subsection{Optimization and Tricks}\label{optimization}
We summarize some optimizations and technique tricks in FB$^+$-tree:

\vspace{0.3em}
\textit{\textbf{Optimization.}} As discussed in previous work, tree nodes of four cache lines exhibit the highest overall performance \cite{masstree, openbw}. Modern machines have nearly identical latency for 64-byte and 256-byte memory accesses. For better single-threaded performance and space efficiency, the \textit{ns} and \textit{fs} are configured to 64 and 4, respectively. A smaller fanout increases the tree depth, while a larger feature size entails more bandwidth requirements and incurs higher access latency. A feature size of four ensures that feature comparison would not lead to excessive overhead when it fails to determine the branch.
To minimize space consumption, the inner nodes only store pointers to anchor keys (i.e., \textit{high\_key} in leaf nodes). The \textit{high\_key} serves as the upper bound of a leaf node and can be constructed using discriminative prefixes to improve performance and space consumption. Additionally, anchor keys in an inner node can also be copied to contiguous memory to enhance locality. \par

\vspace{0.3em}
\textit{\textbf{Tricks.}} The relative magnitude between two unsigned integers can be determined by byte-wise comparison. For signed integers, however, such a pattern doesn't work because of its complement representation. In essence, complement representation is designed to utilize the overflow bit to operate positive and negative integers uniformly. The relative magnitude among positive or negative integers depends on the remaining bits except for the signed bit. For example, in 8-bit complement representation, -2, -1, 0, 1, and 2 are represented as 0xFE, 0xFF, 0x00, 0x01, and 0x02, respectively. An unsigned comparison on the integers whose sign bit is flipped can be treated equivalently to a signed comparison on the original integers, and vice versa. Except for AVX512, the unsigned byte comparison instructions are not supported in AVX2 and SSE2. Therefore, we add a magic number 128 before feature comparison in Figure \ref{lst:lookup}, making FB$^+$-tree widely applicable. \par

\begin{figure}[t]
    \centering
    \includegraphics[width=0.4\textwidth]{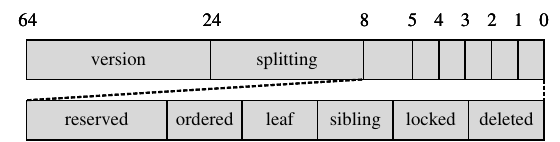}
    \caption{The layout of optimistic lock variable (\textit{control}).}
    \label{fig:control}
    \vspace{-1em}
\end{figure}

\section{synchronization protocol}
\label{section4}
Besides cache utilization, the performance of a main-memory index significantly depends on its synchronization protocol. In this section, we begin by introducing lock-based synchronization protocols and their optimistic variants. Next, we present how FB$^+$-tree is synchronized using an optimistic protocol and an optimization that mitigates the overhead caused by synchronization protocol during index traversal. In particular, we propose a general latch-free update technique that allows updates without holding any locks.

\subsection{Preliminaries to Index Synchronization}\label{sync-primer}
\textcolor{revisioncolor}{Index synchronization consists of two parts: node protection and concurrent structure modification. Traditional database systems typically use pessimistic lock for node protection, along with lock coupling \cite{lock-coupling} or the link technique from B$^{link}$-tree \cite{blink, symmetric} for concurrent structure modification. In main-memory scenarios, optimistic lock combines version with lock to replace pessimistic lock, allowing index traversal without holding locks \cite{cscc, olc, rowex, masstree}. For structure modification, optimistic lock coupling (OLC) keeps track of versions across multiple nodes and restarts if a change occurs \cite{olc, rowex}. OLFIT proceeds to sibling node when detecting a structure modification through comparison with high key \cite{cscc, blink}.}

\textcolor{revisioncolor}{Optimistic lock offers a general approach to accessing a node without holding the lock, making it particularly beneficial for read-heavy workloads. However, update-heavy workloads under contention suffer from performance collapse in two aspects. First, existing optimistic locks are typically implemented as spin locks with optimistic reads \cite{rowex, olc, optik, optiql}. Writers must acquire an exclusive write lock via CAS instruction before modifying a node. This often leads to a scenario where many writers frequently retry CAS in a single node until they hold the lock. A backoff algorithm may somewhat mitigate this collapse, but not completely. Second, readers usually have to wait for the writer's modification and may have to restart if the version validation fails.}

\textcolor{revisioncolor}{OptiQL extended the classic MCS lock with optimistic reads to alleviate the former problem \cite{mcs-lock, optiql}. ROWEX mitigates the latter problem by only providing exclusion relative among writers while allowing readers to always succeed without block nor restart \cite{olc, rowex}. The Bw-tree, incorporating lock-free semantics through chaining delta records and mapping table, seems to be an ideal solution. Unfortunately, delta records require expensive merge operations and the mapping table incurs additional overhead for other operations. FB$^+$-tree alleviates the former problem by minimizing the critical section to allow latch-free update operation. Read operations thus can always succeed when concurrent with updates, which eliminates the latter problem. We show their comparisons when reads are concurrent with updates in Table \ref{tab:sync}. Next, we introduce FB$^+$-tree's synchronization protocol.}

\begin{table}[h]
\vspace{-0.5em}
\renewcommand{\arraystretch}{1.2}
\centering
\caption{Comparisons of Synchronization Protocols}
\resizebox{0.48\textwidth}{!}{
\begin{tabular}{cccccc}
    \toprule
    & OLC & ROWEX & OptiQL & Bw-tree & FB$^+$-tree \\
    \midrule
    latch-free update & & & & \checkmark & \checkmark \\
    no merge overhead & \checkmark & \checkmark & \checkmark & & \checkmark \\
    non-blocking read & & \checkmark & & \checkmark & \checkmark \\
    no auxiliary struct & \checkmark & \checkmark & \checkmark & & \checkmark \\
    \bottomrule
\end{tabular}
}
\label{tab:sync}
\vspace{-0.5em}
\end{table}

\subsection{FB$^+$-tree's Synchronization Protocol}


FB$^+$-tree utilizes a similar optimistic lock for node protection and latch-free index traversal. Figure \ref{fig:control} illustrates the layout of per-node optimistic lock variable, denoted as \textit{control} in Figure \ref{fig:structure}. Insert and remove operations increment the \textit{version}, while update operations do not, which differs from previous work. The \textit{splitting} is only used in leaf nodes, indicating whether the node is undergoing a split and if the new node has not been inserted into its parent node. The \textit{ordered} field indicates whether the key-value pairs in leaf nodes are ordered, which is lazily set only when either split and merge of leaf nodes or range iteration. The \textit{leaf} specifies the type of node. The \textit{sibling} indicates whether the node has a sibling node. The \textit{locked} is set when acquiring an exclusive write lock. The \textit{deleted} is set if the node's contents have been merged into its left-sibling node, indicating that the node can be safely reclaimed later.

\vspace{0.3em}
\textit{Node Protection.} Index traversal loads the \textit{version} before accessing a node and validates that it has not changed after the access. If validation fails, node access needs to restart. Both insert and remove operations acquire an exclusive write lock to prevent concurrent modification within the same node. Since read-only node access may occur concurrently with node modification, atomic operations are employed for insert and remove operations. Lookup operation atomically loads the pointer, thereby preventing readers from accessing a partially modified key-value.

\begin{figure*}[t]
    \centering
    \includegraphics[width=0.95\textwidth]{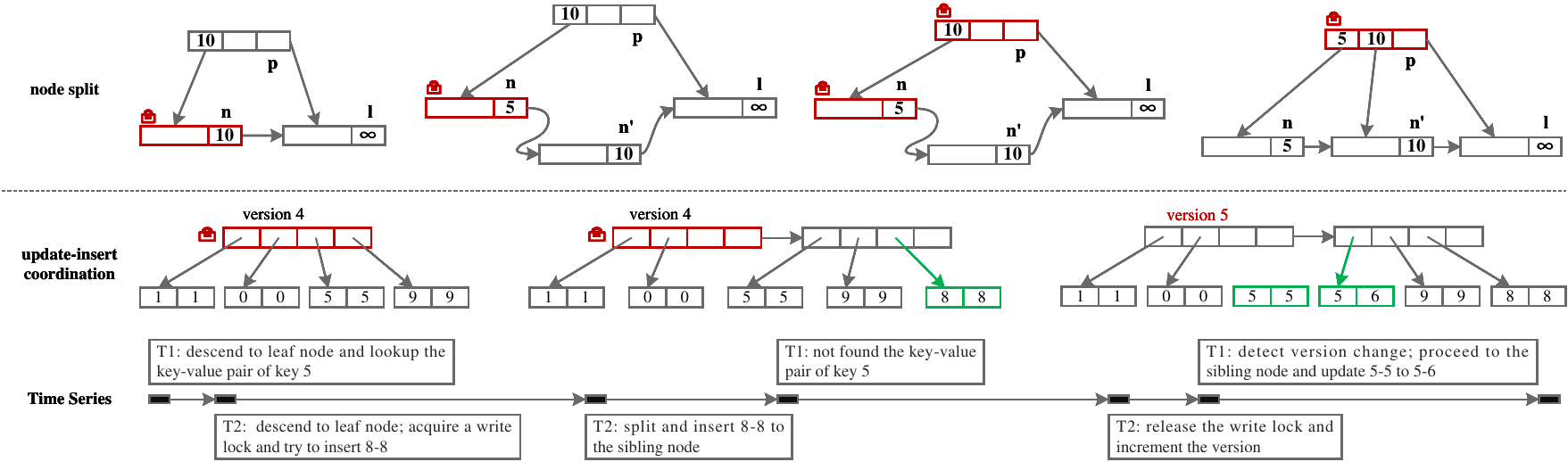}
    \caption{Illustration of structure modification (top half) and update-insert coordination (bottom half, T1-update, T2-insert).}
    \label{fig:synchronization}
    \vspace{-0.5em}
\end{figure*}

\vspace{0.3em}
\textit{Concurrent Structure Modification.} FB$^+$-tree adopts the link technique from B$^{link}$-tree.\footnote{FB$^+$-tree can also utilize lock coupling for concurrent structure modification. We employ the link technique because it facilitates optimizations to latch-free update.} The top half of Figure \ref{fig:synchronization} illustrates the process of concurrent node split.\footnote{Concurrent node merge can be implemented similarly to node split \cite{symmetric}.} A leaf node split involves two steps: \hypertarget{step1}{(1)} transfer the greater half key-value pairs into the newly created node, denoted as \textit{n}$^\prime$; \hypertarget{step2}{(2)} insert the pointer to node \textit{n}$^\prime$ into its parent node \textit{p}. A leaf node \textit{n} is said to be undergoing a split until the pointer to the new node \textit{n}$^\prime$ is inserted into its parent node.

After step \hyperlink{step2}{(2)}, an index traversal could correctly descend to node \textit{n}$^\prime$. An incorrect leaf node occurs only when an index traversal descends to a leaf node that is undergoing a split. Given that the key-values in node \textit{n}$^\prime$ are greater than those in node \textit{n}, it can be addressed through an alternative bypass by linking node \textit{n} to \textit{n}$^\prime$. Therefore, upon descending to a leaf node, a comparison is performed with the upper bound to determine whether it is necessary to proceed to the right sibling node. Split operations that propagate to higher-level nodes can be performed iteratively.

\subsection{Concurrent Lookup}\label{concurrent-lookup}
\begin{figure}[t]
\centering
\Small
\begin{tcolorbox}[colback=white!100,colframe=black!100,boxrule=0.8pt,arc=0pt]
\vspace{-1em}
\begin{lstlisting}[language=C++,keywordstyle=\color{black},morekeywords={String, uint64_t,Control, KVPair, InnerNode, LeafNode},emph={root_},emphstyle={\color{darkgreen}},commentstyle=\color{black!50!white},basicstyle=\fontsize{6.2}{6.2}\ttfamily,columns=flexible,numbers=left,numbersep=5pt,tabsize=2,escapeinside=``]
  KVPair* lookup(String& key) {
    void* node = root_;
    while (!is_leaf(node)) {
      node = InnerNode(node)->branch(key); `\label{con:branch}`
    }
    uint64_t version;
    do {  // descending to leaf node
      version = Control(node)->begin_read();
      while (LeafNode(node)->to_sibling(key, node)) { `\label{con:tosib}`
        version = Control(node)->begin_read();
      }  // check whether to proceed to sibling node
      KVPair* kv = LeafNode(node)->lookup(key);  `\label{con:lookup}`
      if (kv != nullptr) return kv; // key found  `\label{con:rkv}`
    } while (!Control(node)->end_read(version));
    return nullptr;  // key not found
}
\end{lstlisting}
\vspace{-1em}
\end{tcolorbox}
\captionsetup{skip=5pt}
\caption{Concurrent Lookup Algorithm}\label{lst:con-lookup}
\vspace{-0.5em}
\end{figure}

\noindent
Figure \ref{lst:con-lookup} presents the concurrent lookup algorithm. The \textit{branch} algorithm (line \ref{con:branch}) has been slightly modified to start by loading the \textit{version} using \textit{begin\_read} and to restart if validation fails using \textit{end\_read}. Unlike its sequential version in Figure \ref{lst:lookup}, the \textit{branch} may return a sibling node if a node split occurs. Similarly, the \textit{lookup} in leaf node (line \ref{con:lookup}) is protected by \textit{begin\_read} and \textit{end\_read}. To avoid retrieving an incorrect leaf node, the \textit{to\_sibling} function (line \ref{con:tosib}) performs a comparison with \textit{high\_key} upon descending to a leaf node. If the key-value pair is found, the result is returned directly to eliminate unnecessary restarts (line \ref{con:rkv}). 

\vspace{0.3em}
\textit{Cross-Node Tracking.} The overhead of comparison with \textit{high\_key} is negligible for traditional systems, but it may be expensive in main-memory scenarios. Since structure modifications occur infrequently, this comparison is unnecessary in most cases. We eliminate this comparison overhead through a technique named cross-node tracking. Index traversal descends to an incorrect leaf node only when the node is undergoing a split. Additionally, a leaf node split ends up after inserting the new anchor key into its parent node.

We thereby embed a \textit{splitting} field in \textit{control} indicating the node is undergoing a split, which is set when step \hyperlink{step1}{(1)} begins. Once the new anchor key is inserted into the parent node in step \hyperlink{step2}{(2)}, the \textit{splitting} field is unset and the \textit{version} of the parent node is incremented. Through validating the \textit{splitting} and the \textit{version} of the parent node, it can be shown that index traversal descends to a correct leaf node. Consequently, the comparison is performed only when the \textit{splitting} is set or the \textit{version} of the parent node has changed.

\begin{figure}[t]
\centering
\small
\begin{tcolorbox}[colback=white!100,colframe=black!100,boxrule=0.8pt,arc=0pt]
\vspace{-1em}
\begin{multicols}{2}{}
\begin{lstlisting}[language=C++,basicstyle=\fontsize{6.2}{6.2}\ttfamily,columns=flexible,numbers=left,numbersep=5pt,tabsize=2,escapeinside=``,keywordstyle=\color{black},morekeywords={},commentstyle=\color{black!50!white}]
`\colorbox{lightgray!20}{\parbox{0.4\textwidth}{\raggedright node = index\_traversal()}}`
`\colorbox{red!15}{\parbox{0.4\textwidth}{\raggedright node->lock\_exclusive()}}`
`\colorbox{red!15}{\parbox{0.4\textwidth}{\raggedright ... querying / validation}}`
`\colorbox{red!15}{\parbox{0.4\textwidth}{\raggedright ... install kv into kvs}}`
`\colorbox{red!15}{\parbox{0.4\textwidth}{\raggedright node->unlock\_exclusive()}}`
\end{lstlisting}
\begin{lstlisting}[language=C++,basicstyle=\fontsize{6}{6.5}\ttfamily,columns=fixed,numbers=left,numbersep=5pt,tabsize=2,escapeinside=``,keywordstyle=\color{darkgreen},morekeywords={},commentstyle=\color{black!50!white}]
`\colorbox{lightgray!20}{\parbox{0.4\textwidth}{\raggedright node = index\_traversal()}}`
`\colorbox{lightgray!20}{\parbox{0.4\textwidth}{\raggedright ver = node->begin\_read()}}`
`\colorbox{lightgray!20}{\parbox{0.4\textwidth}{\raggedright ... querying / validation}}`
`\colorbox{red!15}{\parbox{0.4\textwidth}{\raggedright ... install kv into kvs}}`
`\colorbox{lightgray!20}{\parbox{0.4\textwidth}{\raggedright node->end\_read(ver)}}`
\end{lstlisting}
\end{multicols}
\vspace{-1em}
\end{tcolorbox}
\captionsetup{skip=5pt}
\caption{Lock-based (left) and latch-free update (right).}
\label{fig:critical-section}
\vspace{-0.5em}
\end{figure}

\subsection{Latch-free Update}\label{lf-update}

\noindent
We propose a general latch-free update technique to achieve good scalability in heavy-contention scenarios. Figure \ref{fig:critical-section} with the critical sections highlighted illustrates the key difference between the traditional lock-based approach and latch-free update. One primary reason for their sub-optimal scalability is that too many ancillary operations are protected inside the critical section. In previous optimistic protocols, a write lock is held to prevent other concurrent modifications when reaching the leaf node. Unfortunately, querying the node for the existing key-value pairs or validation process is also protected in the critical section. Our latch-free update allows this process to be executed in parallel. Only the process of installing kv into \textit{kvs} is protected in the critical section.

\textcolor{revisioncolor}{When only considering update and lookup, tree structure and key-value residences in leaf nodes never change. Update operation thus can be easily implemented using CAS primitive. The challenging problem arises when another thread concurrently performs a split operation on the leaf node. As a result, the key-value to be replaced might have been moved to another node. For example, consider an update operation that tries to replace the key-value 5-5 with 5-6 as shown in the bottom half of Figure \ref{fig:synchronization}. In this case, the update descends to the leaf node and then stalls, while another thread performs a node split on this node. When the update resumes, it definitely fails because the key-value with key 5 cannot be found. Similar issues exist in node merge and node rearrangement.}

\textcolor{revisioncolor}{Our latch-free update solves this false negative problem by checking the version. If the version has not changed, the update fails because the key-value to be updated has not been inserted yet. Otherwise, if the version changes, it implies that the key-value may have been moved to another node. In such cases, a comparison with \textit{high\_key} is then performed to determine whether the key-value is beyond the current node. If so, it proceeds to the sibling node to perform the update again, as illustrated in the bottom half of Figure \ref{fig:synchronization}. Otherwise, it indicates that either the node may have been rearranged or the key-value may have been removed. Restart in the leaf node could deal with these situations.}

To coordinate with latch-free update when moving key-value pairs to another node, we utilize atomic exchange instruction to replace the pointer with \textsc{nullptr} and obtain the latest pointer to a key-value. Concurrent updates would thus fail because they load a \textsc{nullptr} or fail to perform CAS, and then proceed to the sibling node accordingly. Node merge and rearrangement coordinate with updates similarly. Consequently, updates are performed in an almost non-blocking way, and lookups can be executed concurrently.

\subsection{Range Iteration} \label{range-iteration}
In FB$^+$-tree, all the leaf nodes are linked in a totally ordered list. A concurrent scan operation can be performed in two steps: \hypertarget{range1}{(1)} find the starting point on the leaf node list; \hypertarget{range2}{(2)} sequentially iterate on the list. The former is achieved using the \textit{upper\_bound} and \textit{lower\_bound} functions, and the latter is implemented with a concurrent iterator.

\vspace{0.3em}
\textit{Lazy Rearrangement.} Step \hyperlink{range1}{(1)} is akin to a lookup operation except lazy rearrangement. Since maintaining key-values in order is expensive for both lookup and insert, FB$^+$-tree stores unsorted key-values in leaf nodes. The \textit{ordered} bit is embedded in \textit{control} to indicate whether the key-values are in order. When descending to a leaf node, it checks the \textit{ordered} bit. If unordered, it acquires a write lock and then rearranges the key-value pointers, allowing range iteration to benefit from a sequential memory access pattern. Otherwise, it finds the start point without holding the lock. It should be noted that lazy rearrangement incurs a small overhead, because over half of key-values are sorted during node split or merge.

FB$^+$-tree's concurrent iterator in step \hyperlink{range2}{(2)} is almost identical to an STL iterator. In concurrent environments, however, an iterator has to coordinate with insert and remove operations. FB$^+$-tree's iterator thus contains a version to detect whether any modifications occur in a leaf node. If the version changes during iteration, the \textit{ordered} would be checked and the node could be rearranged when necessary. The iterator can thus access the newly inserted key-values. 

\vspace{0.3em}
\textit{Cross-Node Tracking.} If a node split occurs when the iteration crosses leaf nodes, the successor key-value is determined using binary search. The cross-node tracking technique is employed to detect if any structure modifications occur when crossing nodes during range iteration. If no changes have occurred after proceeding to the sibling node, the iteration could optimistically access the minimum key-value for better performance.


\section{experiment evaluation}
\label{section5}
In this section, we experimentally evaluate FB$^+$-tree and compare it with other state-of-the-art main-memory index structures.

\subsection{Experiment Setup}\label{sec:exp-set}
\paragraph{\textbf{Platform}} We use a Dell PowerEdge R740 Server with two NUMA nodes. Each node contains an Intel Xeon(R) Gold 6248R processor with 24 3.0 GHz cores (with up to 48 hyperthreads). Each processor has 35.75 MB L3 cache and is equipped with 64 GB DDR4-2133 memory. We run Ubuntu 20.04 with kernel version 5.4.0. All our code is implemented with C++ 17 and compiled using GCC/G++ 11.4.0 with O3 optimization level. We use jemalloc to reduce dynamic memory allocation overhead at runtime. Threads are pinned to hardware threads to avoid migrations by the OS scheduler.

\paragraph{\textbf{Indexes}} We compare FB$^+$-tree with seven popular main-memory index structures, including the variants of both B$^+$-tree and trie:
\begin{itemize}[leftmargin=*]
    \item STX B$^+$-tree\footnote{\url{https://github.com/tlx/tlx.git}}: A highly optimized B$^+$-tree container with improved memory fragmentation and cache efficiency \cite{TLX}.

    \item FAST\footnote{\url{https://github.com/RyanMarcus/fast64.git}}: \textcolor{revisioncolor}{A read-only binary search tree that collapses multiple nodes into one large node to facilitate SIMD instructions \cite{fast}.}
    
    \item B$^+$-treeOLC\footnote{\url{https://github.com/wangziqi2016/index-microbench.git}\label{index-olc}}: A thread-safe B$^+$-tree implementation synchronized by optimistic lock coupling \cite{openbw}.
    
    \item ART: The default index of HyPer \cite{hyper}. We use two thread-safe implementations ARTOLC\textsuperscript{\ref{index-olc}} and ARTOptiQL\footnote{\url{https://github.com/sfu-dis/optiql}}, synchronized by optimistic lock coupling and OptiQL, respectively \cite{openbw, optiql}.
    
    \item HOT\footnote{\url{https://github.com/speedskater/hot.git}}: The Height Optimized Trie dynamically varies the number of discriminative bits considered in each node\cite{hot}. HOT utilizes the ROWEX protocol for synchronization \cite{rowex, olc}.

    \item Masstree\footnote{\url{https://github.com/kohler/masstree-beta.git}}: The Masstree is a trie-like concatenation of B$^+$-tree used by silo \cite{masstree, silo}. It employs a customized optimistic lock protocol along with the link technique for synchronization.

    \item Wormhole\footnote{\url{https://github.com/wuxb45/wormhole.git}}: The Wormhole substitutes inner nodes of B$^+$-tree with a trie structure and represents it as a hash table. It takes $O(log \ L)$ worst-case time for querying a key of length $L$ \cite{wormhole}.
\end{itemize}
In STX B$^+$-tree, B$^+$-treeOLC, and FB$^+$-tree, integer keys are stored directly in inner nodes, whereas string keys are stored via pointers. In our experiments, all indexes maintain a pointer to each key-value. To avoid excessive compiler optimization, such as unused result optimization, we compile the code of index structures into a shared library. For FB$^+$-tree, the \textit{ns} and \textit{fs} are configured to 64 and 4 respectively, and AVX2 instructions are configured as default. Other indexes are configured to their default configurations. We do not compare against learned indexes, as these hardly support insert operation. We also do not compare against hash tables, because these do not support range iteration.

\paragraph{\textbf{Workloads}} Our experiments are based on the standard workloads from the Yahoo! Cloud Serving Benchmark\footnote{\url{https://github.com/brianfrankcooper/YCSB/}} (YCSB) \cite{ycsb}. We evaluate four core workloads with the default YCSB parameters \textcolor{revisioncolor}{(requests follow the Zipfian distribution, skew=0.99)}: LOAD (100\% insert), A (50\% read, 50\% update), C (100\% read), and E (95\% range-scan, 5\% insert). \textcolor{revisioncolor}{Our main concern is the index structure itself, so range scan only reads pointers without actually accessing the records.} Each workload consists of two phases: the load phase inserts 100 million keys in random order into indexes (one percent of keys are inserted for warmup); the run phase executes 100 million operations specified by the workload multiple times. We evaluate each workload on five datasets with different key lengths (Table \ref{tab:datasets}). The \textit{rand-int} consists of random 64-bit integers. The \textit{3-gram}\footnote{\url{https://www.statmt.org/lm-benchmark/}} \cite{ngrams} contains unique sequences of triple words often used in language models. We also use the default keys generated by YCSB. The \textit{twitter}\footnote{\url{https://github.com/twitter/cache-trace}} (cluster-27) is anonymized data by collecting real-world production traces from in-memory cache clusters at Twitter \cite{twitter}. The \textit{url} consists of distinct URLs in DBPedia dataset \cite{dbpedia}.

\begin{table}[H]
\centering
\caption{Datasets used in the experiments.}
\resizebox{0.45\textwidth}{!}{
\begin{tabular}{ccc}
    \toprule
    Name & Description & Avg. key size (bytes) \\
    \midrule
    rand-int & 64-bit random integers & 8.0 \\
    3-gram & unique sequences of triple words & 15.8 \\
    ycsb & keys generated by YCSB & 22.9 \\
    twitter & anonymized keys in cluster-27 & 52.7 \\
    url & urls in DBPedia dataset & 76.8 \\
    \bottomrule
\end{tabular}
}
\label{tab:datasets}
\end{table}

\subsection{Comparison against B$^+$-tree Variants}
\begin{figure}
    \centering
    \includegraphics[width=0.48\textwidth]{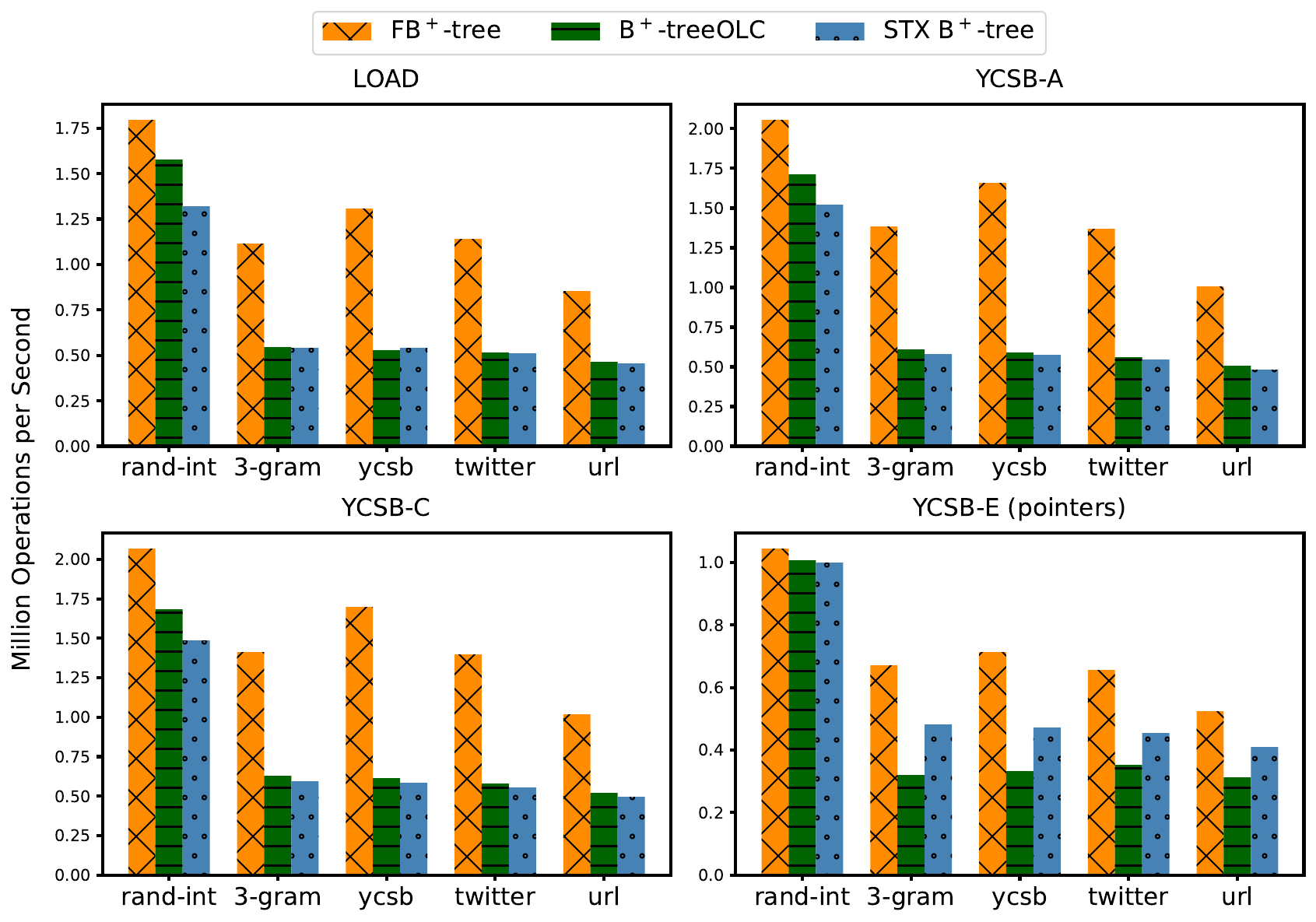}
    \captionsetup{skip=5pt}
    \caption{Single-threaded throughput of B$^+$-tree variants.}
    \label{fig:btree-ycsb}
    \vspace{-0.5em}
\end{figure}
We first measure the single-threaded performance of B$^+$-tree variants to evaluate the structural optimization of FB$^+$-tree. Figure \ref{fig:btree-ycsb} shows the throughput of FB$^+$-tree and two competitors, STX B$^+$-tree and B$^+$-treeOLC. On all workload-dataset combinations, FB$^+$-tree outperforms the other two competitors\textemdash by up to 2.5x (LOAD), 2.9x (YCSB-A), 2.9x (YCSB-C), and 2.2x (YCSB-E). As presented in Section \ref{feature-cmp}, the feature comparison technique significantly reduces cache misses and enables memory-level parallelism, leading to superior single-threaded lookup, update, and scan performance. Besides feature comparison, the unordered arrangement of key-values in FB$^+$-tree contributes to efficient insert performance. \textcolor{revisioncolor}{Since FAST is a read-only structure and only supports integer keys, we compare it to FB$^+$-tree in a separate evaluation, as shown in Figure \ref{fig:cache_miss}.}

\subsection{Evaluation on Our Optimizations}\label{sec:eva-opts}
\begin{figure}
    \centering
    \includegraphics[width=0.48\textwidth]{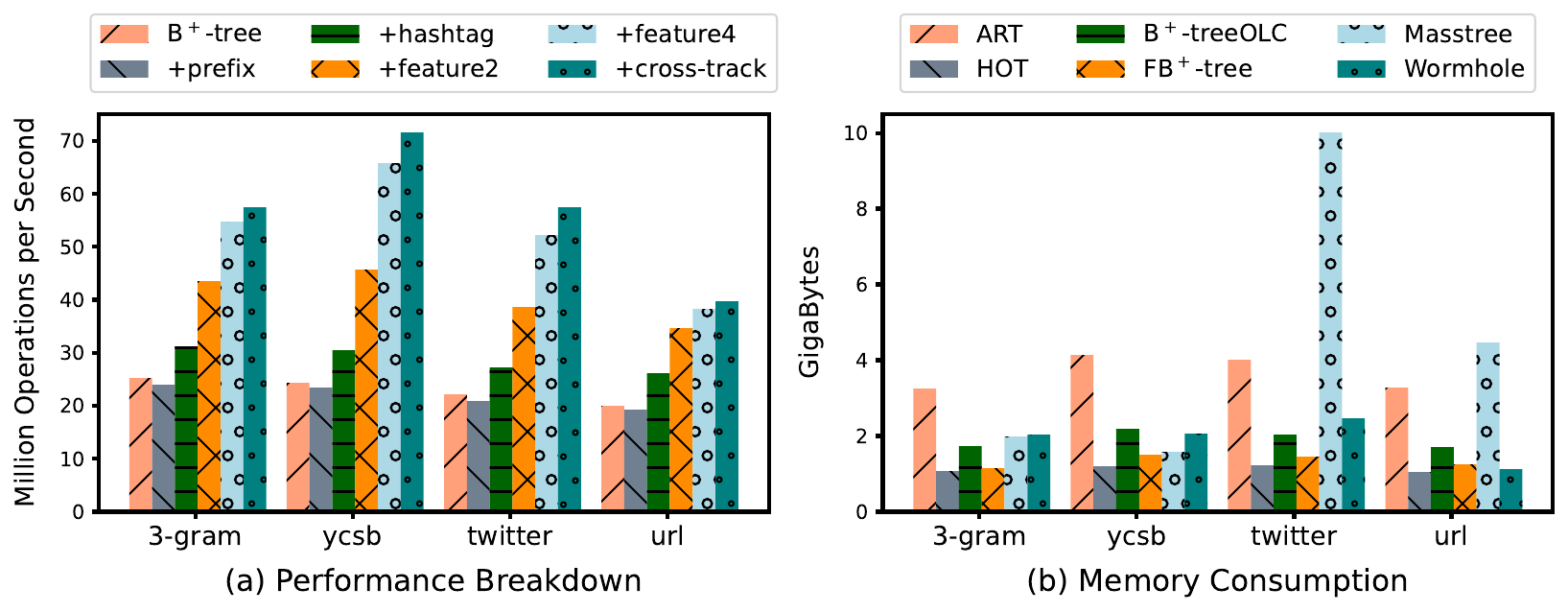}
    \captionsetup{skip=5pt}
    \caption{(a) FB$^+$-tree's multi-threaded (48 cores) throughput on workload YCSB-C by gradually enabling the optimizations, and (b) Memory consumption of different indexes.}
    \label{fig:break-foot}
    \vspace{-0.5em}
\end{figure}

\textbf{\textit{Factor Analysis on Structural Optimizations.}} We gradually enable the optimizations to evaluate the multi-threaded throughput of FB$^+$-tree, as shown in Figure \ref{fig:break-foot}(a). We initially consider the B$^+$-tree without any optimization, which employs binary search in inner and leaf nodes. Next, we enable prefix (denoted as +prefix) and hashtag (in leaf nodes, called +hashtag) in sequence to evaluate the structural optimizations described in Section \ref{node-imp}. The +prefix only stores the common prefixes of anchors directly in inner nodes, in which branch operation compares the target key with the common prefix and then performs a binary search on the suffix. 

It even decreases the performance, since more cache lines need to be loaded, as discussed in Section \ref{dissonance}. We then evaluate the effects of the feature comparison technique (i.e., +feature2 and +feature4, by configuring two and four features, respectively), which improves cache utilization and exploits memory-level parallelism. Lastly, we enable the cross-node tracking optimization (denoted as +cross-track), which eliminates the overhead of accessing the \textit{high\_key} in leaf nodes, as detailed in Section \ref{concurrent-lookup}. Since some optimizations in Figure \ref{fig:break-foot}(a) are not employed for binary keys, we do not evaluate these on rand-int dataset. In conclusion, the multi-threaded (48 cores) throughput of FB$^+$-tree (+cross-track) on YCSB-C is higher than that of typical B$^+$-tree\textemdash by 2.3x (\textit{3-gram}), 2.9x (\textit{ycsb}), 2.6x (\textit{twitter}), 2.0x (\textit{url}), and 2.1x (\textit{rand-int}, as shown in Figure \ref{fig:indexes-ycsb}).

\begin{figure}[t]
    \centering
    \includegraphics[width=0.48\textwidth]{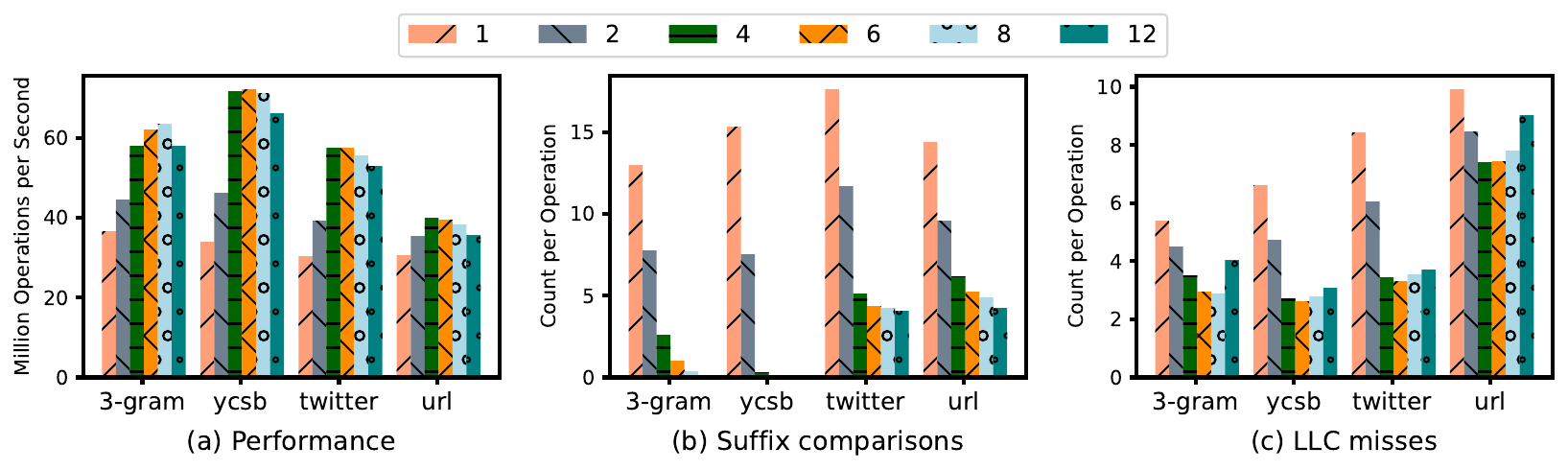}
    \captionsetup{skip=5pt}
    \caption{\textcolor{revisioncolor}{Evaluation with different feature size (YCSB-C).}}
    \label{fig:dif-fsize}
    \vspace{-0.5em}
\end{figure}

\textcolor{revisioncolor}{To deeply evaluate the impact of feature size, we also evaluate the multi-threaded (48 cores) performance, average suffix comparison count per operation, and average LLC-misses per operation with different feature sizes, as shown in Figure \ref{fig:dif-fsize}. Since the feature size of binary keys is fixed, we do not evaluate these on rand-int dataset. Although the average suffix comparison count per operation gradually declines as feature size increases, the average LLC-misses per operation first decreases and then gradually increases, as shown in Figure \ref{fig:dif-fsize}(b)(c). This occurs because the mechanical selection of anchor keys is unconscious of prefix skewness, as discussed in Section \ref{feature-cmp}. As a result, the performance of FB$^+$-tree first increases with feature size, then gradually decreases, as shown in Figure \ref{fig:dif-fsize}(a). In addition, Figure \ref{fig:dif-fsize} also illustrates the reason why FB$^+$-tree's performance varies across different datasets. The twitter and url datasets have more complicated prefix patterns leading to more suffix comparisons and LLC-misses.}

\begin{figure}[t]
    \centering
    \includegraphics[width=0.48\textwidth]{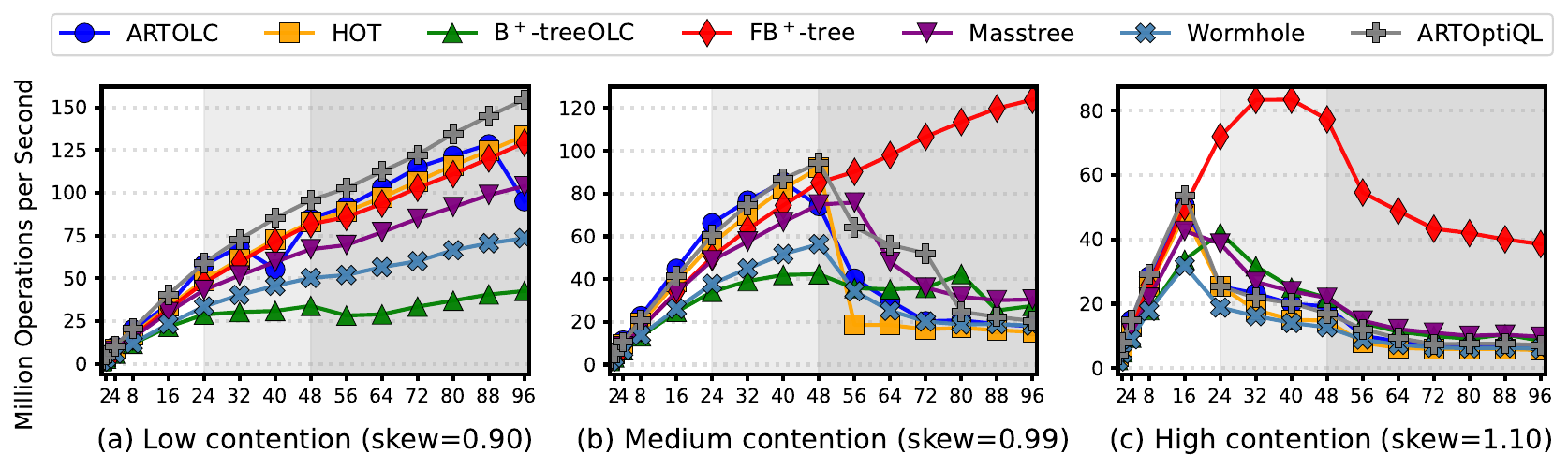}
    \captionsetup{skip=5pt}
    \caption{\textcolor{revisioncolor}{Scalability under different skews (YCSB-A).}}
    \label{fig:skewed-update}
    \vspace{-0.5em}
\end{figure}

\textbf{\textit{Scalability of Latch-free Update.}} \textcolor{revisioncolor}{First, we compare FB$^+$-tree with other state-of-the-art index structures to evaluate their scalability on rand-int dataset using YCSB-A (update-heavy) workload with different skews. The results are shown in Figure \ref{fig:skewed-update}. All index structures are multi-core scalable under low contention. Thanks to its latch-free update and non-blocking read described in Section \ref{sync-primer} and Section \ref{lf-update}, FB$^+$-tree is almost linearly scalable under median contention, while other index structures suffer from performance collapse. Due to retrying CAS on the pointers, FB$^+$-tree also experiences performance collapse under high contention. However, it still maintains the best performance thanks to its minimal hardware-level critical section, as illustrated in Figure \ref{fig:skewed-update}(c).}

\begin{figure}[b]
    \centering
    \includegraphics[width=0.48\textwidth]{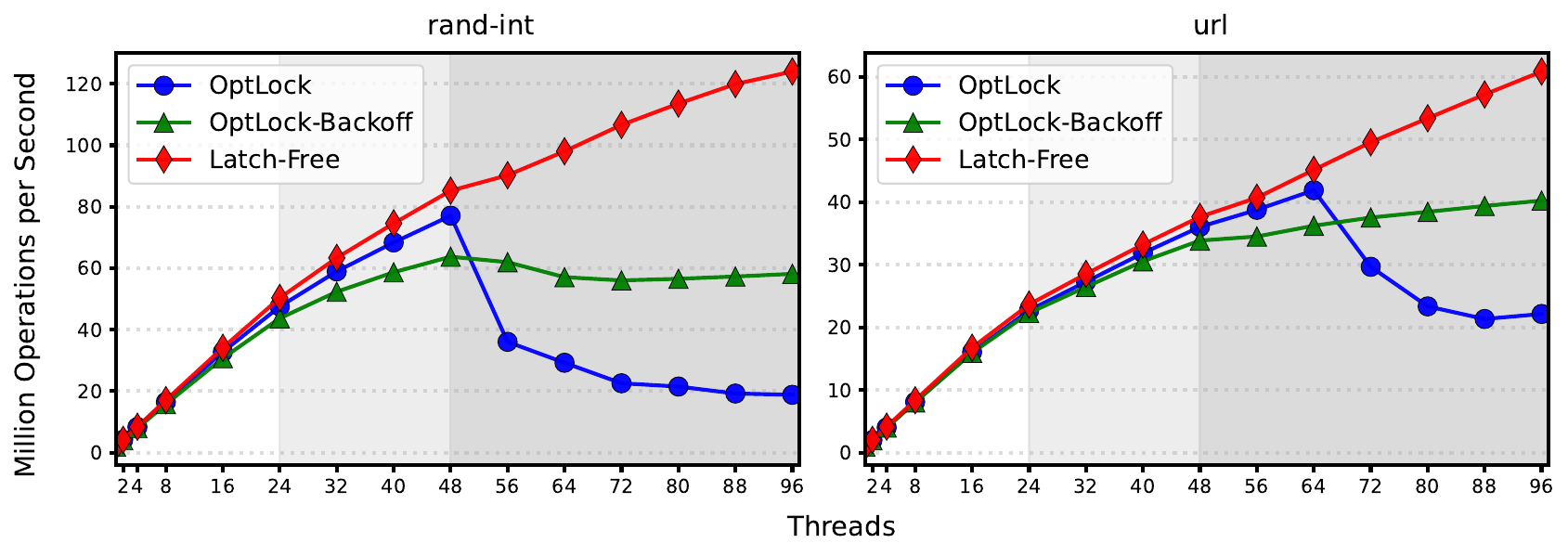}
    \captionsetup{skip=5pt}
    \caption{Scalability on rand-int and url datasets (YCSB-A).}
    \label{fig:lock-vs-latchfree}
    \vspace{-0.5em}
\end{figure}

Next, we horizontally compare the latch-free update with two other competitors, optimistic lock using CAS primitive (denoted as OptLock) and optimistic lock with backoff (called OptLock-Backoff). The scalability of the three techniques on FB$^+$-tree is illustrated in Figure \ref{fig:lock-vs-latchfree}. Due to space limitation, we only evaluate the \textit{rand-int} and \textit{url} datasets (the highest and lowest performance on YCSB-C) with the standard workload YCSB-A (skew=0.99). The OptLock suffers from scalability collapse over 48 threads on \textit{rand-int} dataset and over 64 threads on \textit{url} dataset, respectively. The OptLock with backoff algorithm could mitigate such performance collapse, however, leading to performance degradation with fewer threads. Our latch-free update exhibits the best scalability and outperforms OptLock by 6.6x (\textit{rand-int}) and 2.8x (\textit{url}) under 96 threads.

\subsection{Comparison against State-of-the-art}\label{sec:cmp-sota}
We compare FB$^+$-tree with five state-of-the-art main-memory indexes to evaluate their scalability and performance in a concurrent environment. Meanwhile, we use the statistic interface of jemalloc to evaluate their space efficiency.

\textbf{\textit{Memory Consumption.}} Since these indexes utilize different key-value storage formats, we only report the index memory consumption (the memory required by the index, including the pointers to key-values but excluding the key-values). The Masstree employs a complicated design, in which key-values are not stored together and key slices are stored in its inner nodes. We count the whole memory footprint and then subtract the memory footprint for storing key-values as Masstree's memory consumption. The results on \textit{3-gram}, \textit{ycsb}, \textit{twitter}, and \textit{url} datasets are presented in Figure \ref{fig:break-foot}(b). The results on \textit{rand-int} dataset is consistent with results on \textit{3-gram} dataset. HOT only considers discriminative bits in inner nodes and thus is highly space-efficient. Except for HOT, FB$^+$-tree is more space-efficient than other indexes on almost all datasets.

\begin{figure}[b]
    \centering
    \includegraphics[width=0.48\textwidth]{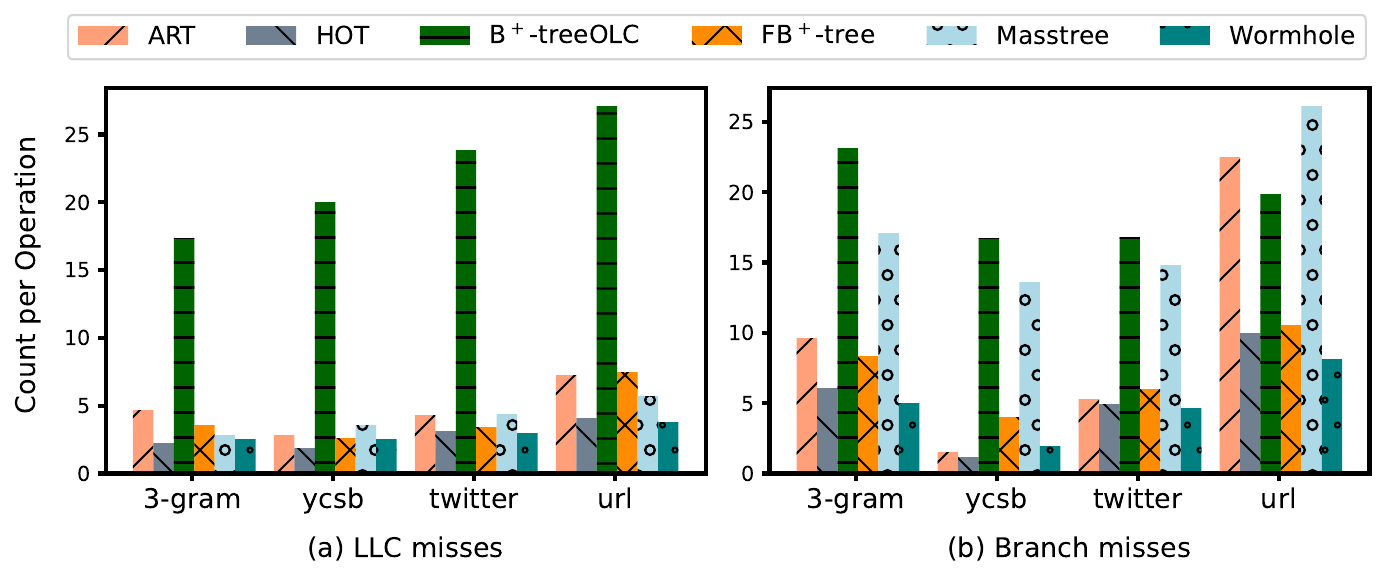}
    \captionsetup{skip=5pt}
    \caption{\textcolor{revisioncolor}{Hardware events count on workload YCSB-C.}}
    \label{fig:statistic}
    \vspace{-0.5em}
\end{figure}

\begin{figure*}
    \centering
    \includegraphics[width=0.98\textwidth]{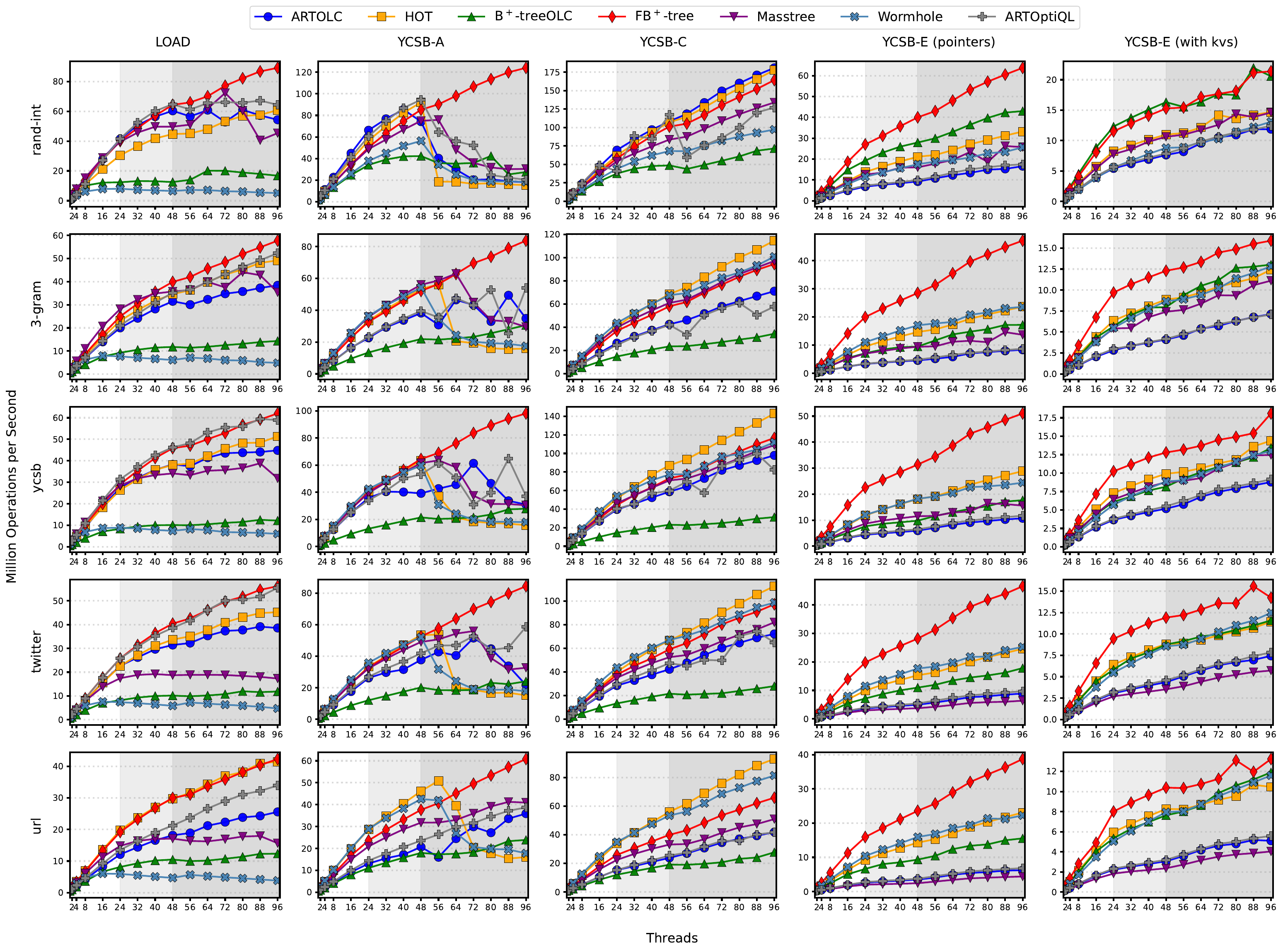}
    \captionsetup{skip=5pt}
    \caption{\textcolor{revisioncolor}{Index throughput and scalability on different workload-dataset combinations.}}
    \label{fig:indexes-ycsb}
    \vspace{-0.5em}
\end{figure*}

\textbf{\textit{Performance and Scalability.}} \textcolor{revisioncolor}{The throughput and scalability of these indexes on all workload-dataset combinations are illustrated in Figure \ref{fig:indexes-ycsb}. On workload LOAD, FB$^+$-tree outperforms all other indexes on all datasets. ARTOptiQL and HOT exhibit comparable scalability and performance on \textit{ycsb}, \textit{twitter}, and \textit{url} dataets. Thanks to the latch-free update technique, FB$^+$-tree shows the best scalability on workload YCSB-A. All other indexes suffer from performance collapse as threads increase. On the read-only workload YCSB-C, HOT gives the best performance and scalability on all datasets except rand-int. FB$^+$-tree performs almost as fast as other trie-based structures. As shown in Figure \ref{fig:statistic}, the average LLC-misses and branch misses per operation of all index structures are counted under a multi-threaded (48-core) environment as a shred of evidence. Except for these two metrics, parallelism between instructions, memory access patterns, and memory bandwidth utilization are also significant, which leads to FB$^+$-tree's better performance than ART and Mastree. It should be mentioned that all these structures perform lookup without holding any locks.}

\textcolor{revisioncolor}{The performance of range scan may be vitally important in many database applications, especially when secondary indexes are extensively utilized to retrieve valuable information and translate database operators into computations based on primary keys. In these scenarios, a balanced structure may be extremely effective. FB$^+$-tree exhibits superior performance than other structures on workload YCSB-E thanks to its balanced structure and sequential pointer arrangement as described in Section \ref{range-iteration}. Although Wormhole has a similar leaf node structure, its indirect ordered key-value arrangement hinders efficient range scan. Frequent pointer chasing in trie-based index structures leads to inferior range scan performance. For completeness, we also illustrate the range scan performance with the actual key-value records access. In summary, the experiments demonstrate that FB$^+$-tree dominates existing B$^+$-tree variants on all the workload-dataset combinations. Compared with trie-based structures, FB$^+$-tree may have lower performance on point lookup, while it has a big advantage on range scan. On update-heavy workloads, FB$^+$-tree also demonstrates significant potential and better scalability over other structures.}

%

\section{conclusion}
\label{section6}
In this paper, we present the FB$^+$-tree, a fast, cache-efficient, and balanced B$^+$-tree variant taking memory-level and computational parallelism into consideration. We show how to reduce cache misses in binary search and exploit prefetching to leverage memory-level parallelism. We highlight our feature comparison technique enabling vectorization of binary search from a different perspective than previous work. The evaluation results demonstrate FB$^+$-tree exhibits comparable point lookup performance to state-of-the-art indexes, while exhibiting superior range scan performance. We sincerely believe that mitigating dependences and exploring the possibility of parallelization and vectorization would be increasingly important to improve the performance of existing algorithms.


\begin{acks}
This research was sponsored by the National Key Research and Development Program of China (No. 2023YFC3321304), the National Science Foundation of China (No. 61572373), and CCF-Huawei Populus Grove Fund (No. CCF-HuaweiDB202410). We thank the anonymous reviewers for their valuable comments. 
\end{acks}

\balance
\clearpage

\clearpage

\bibliographystyle{ACM-Reference-Format}
\bibliography{fbtree}

\end{document}